\documentclass[preprint,12pt]{elsarticle}

\usepackage{tcolorbox}
\usepackage{graphicx}  
\usepackage{multirow}
\usepackage{multicol}
\usepackage{xcolor}
\usepackage{mdframed}
\usepackage{xcolor}
\usepackage{textcomp}
\usepackage{stfloats}
\usepackage{url}
\usepackage{verbatim}
\usepackage{graphicx}
\usepackage[normalem]{ulem}
\usepackage[ruled,vlined]{algorithm2e}
\usepackage{adjustbox}
\usepackage{tabularx}
\usepackage{hyperref}
\usepackage[table]{xcolor}
\usepackage{subcaption}

\usepackage{amssymb}
\usepackage{amsmath}

\journal{Applied Energy}

\begin{document}

\begin{frontmatter}



\title{
Retrieval-Guided Photovoltaic Inventory Estimation from Satellite Imagery for Distribution Grid Planning
}


\author[label1]{Muhao Guo} 
\author[label1]{Lihao Mai}
\author[label2]{Erik Blasch}
\author[label3]{Jafarali Parol}
\author[label3]{Turki Rakan}
\author[label1]{Yang Weng}
\affiliation[label1]{
            organization={Department of Electrical, Computer and Energy Engineering, Arizona State University},
            addressline={Goldwater Center for Science and Engineering, 650 E Tyler Mall}, 
            city={Tempe},
            postcode={85281}, 
            state={AZ},
            country={United States}}
\affiliation[label2]{organization={Air Force Research Laboratory},
            city={Arlington},
            postcode={22203},
            state={VA},
            country={United States}}

\affiliation[label3]{organization={Kuwait Institute for Scientific Research},
            addressline={8WP4+7JH},
            city={Kuwait City},
            postcode={13109},
            country={Kuwait}}


\begin{abstract}
The rapid expansion of distributed rooftop photovoltaic (PV) systems introduces increasing uncertainty in distribution grid planning, hosting capacity assessment, and voltage regulation. Reliable estimation of rooftop PV deployment from satellite imagery is therefore essential for accurate modeling of distributed generation at feeder and service-territory scales. However, conventional computer vision approaches rely on fixed learned representations and globally averaged visual correlations. This makes them sensitive to geographic distribution shifts caused by differences in roof materials, urban morphology, and imaging conditions across regions. 
To address these challenges, this paper proposes Solar Retrieval-Augmented Generation (Solar-RAG), a context-grounded framework for photovoltaic assessment that integrates similarity-based image retrieval with multimodal vision-language reasoning. Instead of producing predictions solely from internal model parameters, the proposed approach retrieves visually similar rooftop scenes with verified annotations and performs comparative reasoning against these examples during inference. This retrieval-guided mechanism provides geographically contextualized references that improve robustness under heterogeneous urban environments without requiring model retraining.
Experiments on a globally distributed satellite imagery dataset covering six continents demonstrate that the proposed framework achieves 95.4\% accuracy for PV presence detection, 86.2\% accuracy for panel quantity estimation, and 80.2\% accuracy for spatial localization. And, the method outperform both conventional deep vision models and standalone vision-language models. Furthermore, feeder-level case studies show that improved PV inventory estimation reduces errors in voltage deviation analysis and hosting capacity assessment. 
The results demonstrate that the proposed method provides a scalable and geographically robust approach for monitoring distributed PV deployment. This enables more reliable integration of remote sensing data into distribution grid planning and distributed energy resource management.
\end{abstract}


\end{frontmatter}


\section{Introduction}
The global transition toward low-carbon energy systems has positioned solar photovoltaic (PV) technology as one of the fastest-growing sources of electricity generation~\cite{haegel2017terawatt,way2022empirically}. Continuous reductions in PV costs, advances in manufacturing, and supportive policy frameworks have enabled rapid deployment across residential, commercial, and utility-scale applications worldwide~\cite{li2025exarnn,li2025external}. As installed capacity moves toward terawatt-scale levels, distributed rooftop PV systems are increasingly changing the operation and planning of modern distribution grids~\cite{hua2023digital}. Unlike centralized monitored generation assets, rooftop PV installations are often deployed by individual customers and may not be fully recorded in utility databases. This results in significant uncertainty in feeder-level PV inventories, which directly affects distribution system modeling, hosting capacity assessment, voltage regulation, protection coordination, and infrastructure investment decisions~\cite{yu2018deepsolar, sun2022insights}. Therefore, reliable estimation of distributed PV deployment across service territories has become a requirement for modern grid planning and distributed energy resource (DER) integration studies~\cite{chen2023remote,mai2025guaranteed}.

For PV inventory estimation, existing monitoring approaches struggle to scale with the rapid and geographically heterogeneous growth of rooftop PV installations. For example, conventional methods rely on manual inspection~\cite{guo2023graph}, field surveys, and specialized diagnostic equipment. But they are costly, labor-intensive, and difficult to apply across large service territories containing millions of buildings. As distributed PV continues to expand across diverse urban environments, these approaches can not maintain accurate PV deployment records. To overcome these limitations, remote sensing techniques using aerial and satellite imagery were introduced for large-scale PV monitoring~\cite{wang2022deepsolar++}. By high-resolution imagery, such approaches enable automated identification of rooftop PV installations across wide geographic areas. This provides an opportunity to construct scalable PV deployment inventories that support distribution grid modeling and planning~\cite{sun2022estimating}.

Recent advances in computer vision have significantly improved the capability of automated PV detection from satellite imagery. Deep learning models such as convolutional neural networks (CNNs) and vision transformers (ViTs) have demonstrated strong performance in solar panel detection~\cite{matusz2025overview}, rooftop segmentation~\cite{wang2023rooftop}, and anomaly identification~\cite{guo2023accurate, ghahremani2025advancements}. However, these methods often rely on fixed learned representations derived from training datasets that may not capture the diversity of rooftop structures and environmental conditions across different geographic regions. As a result, their performance can degrade under domain shifts caused by variations in roof materials, architectural styles, imaging resolution, weather conditions~\cite{dou2024day}, and urban morphology~\cite{zhou2022domain,wang2018deep}. Addressing such variability typically requires retraining models with region-specific labeled datasets. But this increases the costs for obtaining data and  limits scalability for large-scale PV monitoring applications.

In parallel, recent developments in generative artificial intelligence have introduced multimodal large language models and vision–language models (VLMs) capable of semantic reasoning over visual content~\cite{zhang2024vision, zhang2026large}. These models demonstrate strong zero-shot and few-shot generalization abilities and can produce structured explanatory outputs~\cite{guo2025solar,guo2025cross,guo2023msq}, making them attractive for infrastructure monitoring tasks such as satellite-based assessment of distributed energy resources. However, inference in these models still relies primarily on internally learned representations and globally averaged visual priors. When applied to remote sensing scenarios with significant geographic variability, rooftop structures, architectural styles, and environmental conditions can differ substantially across regions, making photovoltaic identification highly dependent on locally relevant visual patterns. Without explicit access to region-specific reference information, VLM-based predictions may exhibit hallucination or misinterpretation of visual features~\cite{lewis2020retrieval}. Consequently, reliable photovoltaic inventory estimation across geographically diverse regions requires an inference framework capable of incorporating contextual visual evidence and dynamically referencing comparable rooftop cases observed under similar geographic and environmental conditions.

\begin{figure}[h]
\centering
\includegraphics[width=1\columnwidth]{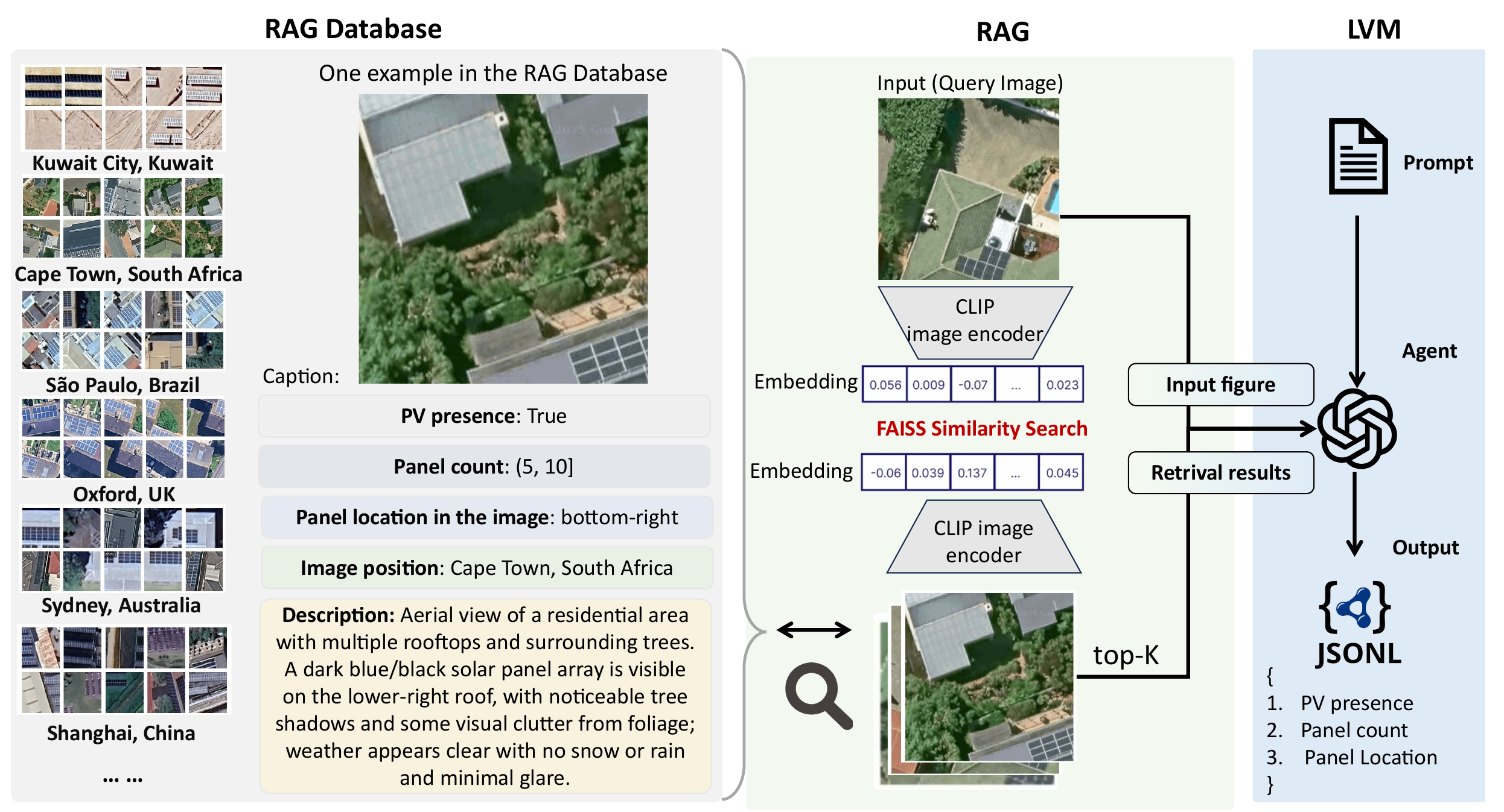}
\caption{Overview of the proposed Solar Retrieval-Augmented Generation (Solar-RAG) framework for photovoltaic assessment from satellite imagery. The framework performs retrieval-guided visual reasoning by combining satellite image embeddings with a globally distributed rooftop reference database. For each query image, visually similar rooftop scenes are retrieved and used as contextual evidence during inference. A vision–language model then performs comparative reasoning between the query rooftop and retrieved reference cases to estimate photovoltaic presence, panel quantity, and spatial location. By incorporating geographically relevant reference information into the inference process, the framework improves robustness under heterogeneous rooftop environments and enables scalable photovoltaic inventory estimation for distribution system analysis.}
\label{fig: Framework}
\end{figure}

To address this challenge, this paper proposes Solar Retrieval-Augmented Generation (Solar-RAG), a context-calibrated framework for photovoltaic assessment from satellite imagery. Unlike conventional pipelines that rely solely on parametric prediction, Solar-RAG performs retrieval-guided visual reasoning by integrating similarity-based rooftop retrieval with multimodal vision–language inference. As illustrated in Fig.~\ref{fig: Framework}, the proposed architecture first identifies visually comparable rooftop scenes from a globally distributed database using visual embedding similarity~\cite{radford2021learning,douze2025faiss}. These retrieved reference examples, together with their structured annotations, are then incorporated into the inference process to enable comparative reasoning over rooftop characteristics. This mechanism allows PV presence, panel quantity, and spatial distribution to be estimated relative to geographically relevant reference cases. Such estimations improve robustness under heterogeneous urban environments and enable scalable PV inventory estimation for distribution system analysis~\cite{yu2024visrag,liu2024remoteclip}.

Unlike conventional pipelines that treat PV detection as a static classification or segmentation task~\cite{zhao2024extracting}, the proposed Solar-RAG performs retrieval-guided and context-aware assessment of rooftop PV installations. The framework evaluates each query image relative to retrieved reference scenes to determine photovoltaic presence, estimate panel quantity, and localize installations within the scene, generating structured outputs suitable for downstream distribution system modeling and techno-economic analysis~\cite{li2025imperfect}. By decoupling knowledge expansion from fixed model parameters and shifting adaptation to a dynamically extendable retrieval layer, Solar-RAG mitigates domain shift, reduces dependence on repeated retraining, and improves interpretability. These capabilities enable scalable PV inventory estimation under geographically heterogeneous conditions, supporting more reliable integration of satellite-derived information into distribution grid planning and DER management~\cite{tiemann2025amplify}.

The rest of this paper is organized as follows. Section~\ref{sec: Problem Formulation} formulates the PV assessment problem. Section~\ref{sec:methodology} presents the proposed methodology. Section~\ref{sec:Experiments} reports experimental results. Section~\ref{sec:ps-coupling} demonstrates feeder-level coupling with AC power-flow analysis. Section~\ref{sec:Conclusion} concludes the paper.

\section{Problem Formulation for PV Inventory Estimation}
\label{sec: Problem Formulation}

Consider a geo-referenced satellite image tile $X$, representing an overhead observation of a built environment within a distribution service territory. The objective of satellite-based photovoltaic (PV) inventory estimation is to extract, from each rooftop observation $X$, a structured set of PV deployment descriptors that can be directly integrated into downstream distribution grid modeling, hosting capacity analysis, and distributed energy resource (DER) planning studies.

For each rooftop image $X$, the assessment output is defined as
\begin{equation}
\mathbf{z}^{\mathrm{PV}} 
=
\left[
p^{\mathrm{PV}}, \,
n^{\mathrm{PV}}, \,
\ell^{\mathrm{PV}}, \,
e^{\mathrm{PV}}
\right],
\end{equation}
where $p^{\mathrm{PV}} \in \{0,1\}$ indicates the presence or absence of rooftop PV installations, $n^{\mathrm{PV}}$ denotes the estimated number of installed panels expressed either as a discrete count or bounded interval, $\ell^{\mathrm{PV}}$ encodes the approximate spatial location of PV arrays within the rooftop image, and $e^{\mathrm{PV}}$ provides a descriptive explanation supporting interpretability, traceability, and engineering validation.

These descriptors provide a compact representation of rooftop PV deployment that can be directly incorporated into distribution system simulations and planning studies. In particular, $p^{\mathrm{PV}}$ and $n^{\mathrm{PV}}$ contribute to installed capacity estimation and distributed generation penetration assessment. The spatial descriptor $\ell^{\mathrm{PV}}$ enables aggregation at feeder, substation, or service-area levels, while $e^{\mathrm{PV}}$ enhances transparency for auditing, regulatory reporting, and operational verification. Given an input rooftop image $X$, the PV inventory estimation task consists of estimating
\begin{equation}
\hat{\mathbf{z}}^{\mathrm{PV}} 
= 
f_{\mathrm{PV}}(X),
\label{eq:pv_mapping}
\end{equation}
where $f_{\mathrm{PV}}(\cdot)$ denotes the estimation mapping from satellite imagery to structured PV deployment descriptors. The mapping in Eq.~\ref{eq:pv_mapping} must operate reliably across heterogeneous geographic regions, architectural typologies, and environmental conditions, while producing standardized outputs compatible with feeder-level modeling, distributed energy resource integration studies, and techno-economic evaluation.
To account for geographic and structural variability in rooftop configurations, the estimation process may incorporate contextual visual information drawn from a reference set of $j = 1, \dots, M$ satellite images
\begin{equation}
\mathcal{K} = \{ X_j^{r} \}_{j=1}^{M},
\end{equation}
where each $X_j^{r}$ represents a rooftop scene exhibiting representative PV deployment patterns under diverse climatic and architectural conditions. The reference set $\mathcal{K}$ provides contextual calibration for estimation under heterogeneous deployment environments, without altering the formal objective defined in Eq.~\ref{eq:pv_mapping}.
Under this formulation, the solar PV inventory estimation problem is defined as estimating $\hat{\mathbf{z}}^{\mathrm{PV}}$ for any rooftop observation $X$, potentially conditioned on contextual information from $\mathcal{K}$, while ensuring scalability, interpretability, and compatibility with distribution system operation and planning workflows.

\section{Solar-RAG Framework for PV Inventory Estimation}
\label{sec:methodology}

This section presents the proposed Solar Retrieval-Augmented Generation (Solar-RAG) framework for satellite-based photovoltaic (PV) inventory estimation. Building upon the formulation introduced in Section~\ref{sec: Problem Formulation}, the framework combines a curated rooftop reference repository with similarity-guided visual reasoning to estimate structured PV deployment descriptors from satellite imagery. Specifically, the approach consists of three components: (i) development of a geographically diverse rooftop reference repository, (ii) similarity-based selection of representative rooftop cases in a structured visual feature space, and (iii) reference-assisted estimation of PV deployment parameters using a multimodal vision–language reasoning module. The resulting PV descriptors $\hat{\mathbf{z}}^{\mathrm{PV}}$ provide structured rooftop information that can be directly incorporated into distribution grid modeling, installed capacity estimation, hosting capacity analysis, and distribution planning studies.

\subsection{Rooftop Reference Repository Construction}
\label{subsec:construction}

\begin{figure}[h]
\centering
\includegraphics[width=0.7\columnwidth]{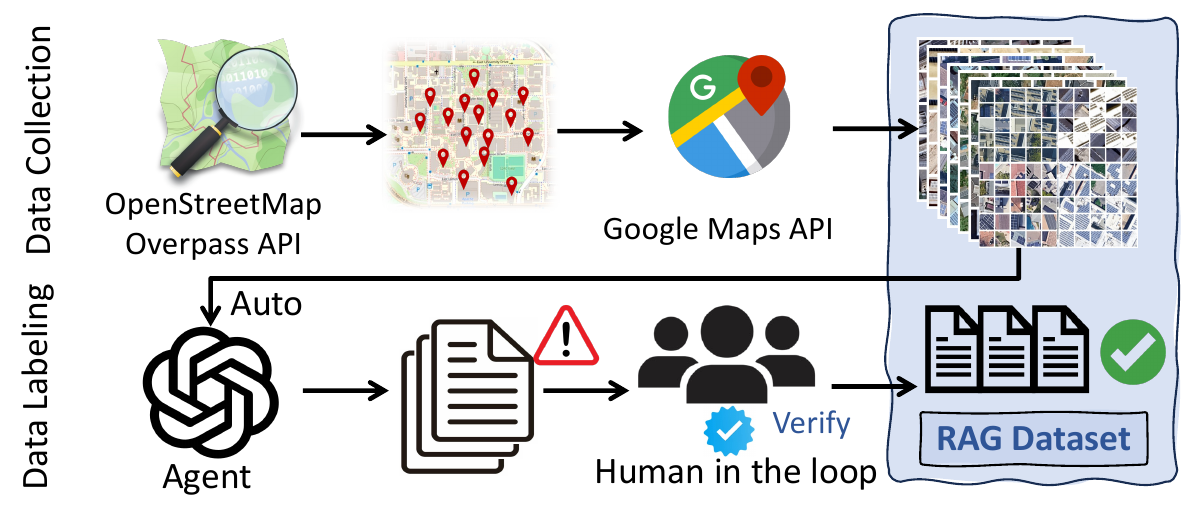}
\caption{Workflow for developing the rooftop reference repository used for distributed PV assessment. Candidate rooftop locations are first identified from open geospatial databases using the OpenStreetMap Overpass API. High-resolution overhead satellite imagery is then retrieved through the Google Maps API under standardized acquisition settings, forming a large rooftop image inventory. An automated attribute extraction module generates preliminary PV descriptors for each tile, including installation presence, panel quantity, spatial location, and contextual characteristics. These descriptors are subsequently reviewed and corrected through human validation to ensure reliability for planning applications. The validated rooftop tiles, together with their structured PV attributes, constitute the reference repository used for similarity-based case selection and reference-assisted PV estimation.}
\label{fig:RAG_dataset}
\end{figure}

The proposed approach relies on a purpose-built rooftop image repository designed to capture diverse PV deployment conditions across geographic regions (see Fig.~\ref{fig:RAG_dataset}). Rather than serving solely as training data for model optimization, this repository functions as an engineering reference library supporting PV estimation under varying climatic, architectural, and urban configurations. Its development follows a structured pipeline integrating geospatial data acquisition, standardized satellite image retrieval, automated semantic extraction, and human validation.

Specifically, geographic coordinates corresponding to residential rooftop locations were obtained from open geospatial repositories using the OpenStreetMap (OSM) Overpass application programming interface (API)\footnote{\url{https://overpass-api.de/}}. These coordinates provide reproducible geographic reference points and explicit documentation of query parameters and coverage areas. High-resolution satellite images were subsequently retrieved through the Google Maps Static API\footnote{\url{https://developers.google.com/maps/documentation/maps-static/overview}} under fixed acquisition settings to ensure consistent rooftop views across cities and to reduce bias associated with scale, zoom level, and orientation.

To balance spatial resolution with scalability, each retrieved satellite image was partitioned into standardized image regions that serve as the fundamental analytical units of the repository. The images were organized by city and further categorized into PV-containing and non-PV subsets according to their contextual information. This structured organization supports balanced similarity-based reference selection during estimation and facilitates systematic expansion as additional regions are incorporated.

\begin{figure}[h]
\centering
\includegraphics[width=1\columnwidth]{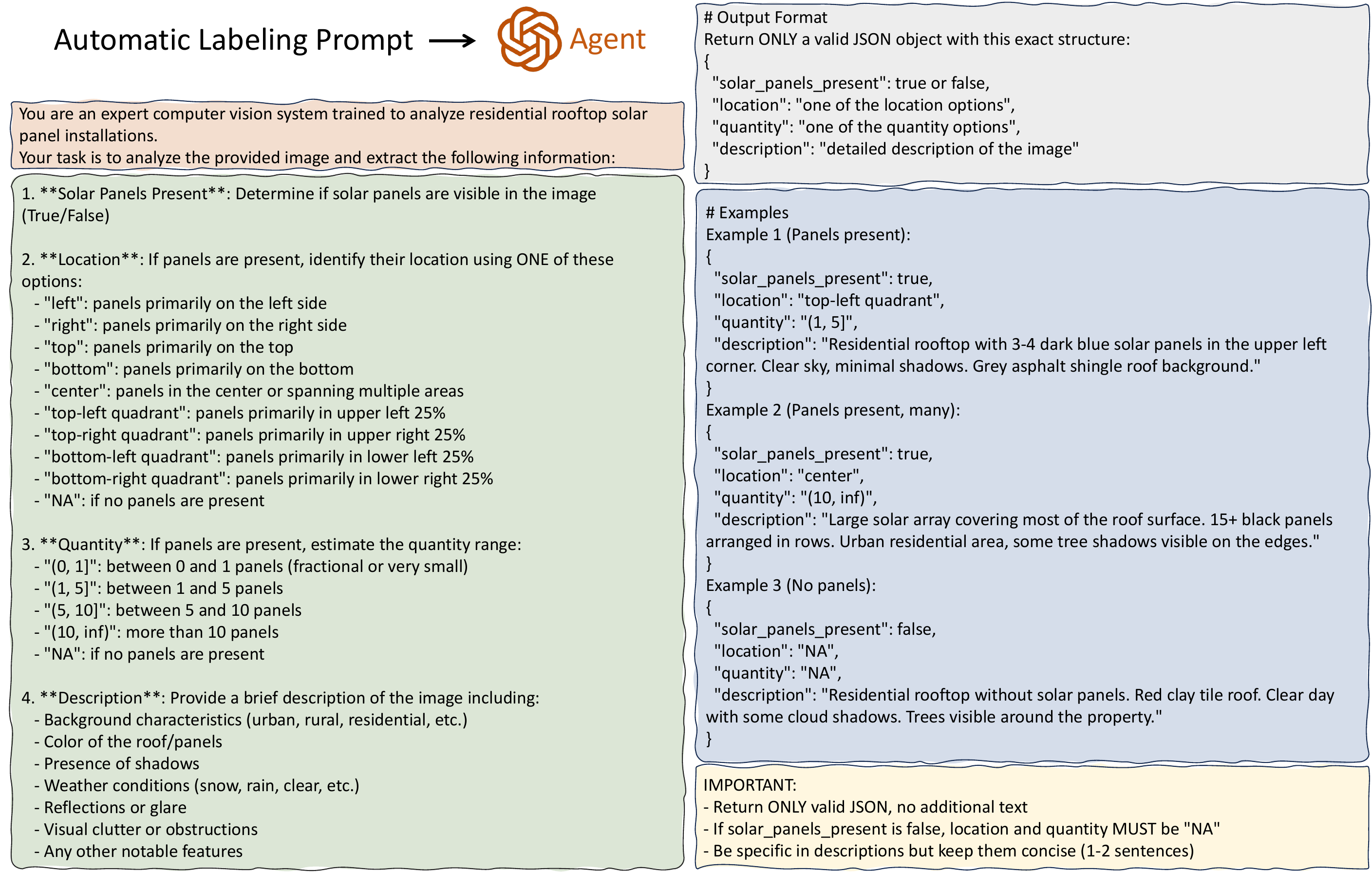}
\caption{Automatic labeling prompt used for structured PV descriptor extraction.}
\label{fig:Automatic_Labeling_Prompt}
\end{figure}

Each satellite image was processed using a structured attribute extraction procedure aligned with the PV descriptor vector $\mathbf{z}^{\mathrm{PV}}$. Specifically, the procedure identifies (i) PV presence $p^{\mathrm{PV}}$, (ii) panel quantity $n^{\mathrm{PV}}$, (iii) approximate spatial location $\ell^{\mathrm{PV}}$, and (iv) descriptive rooftop attributes $e^{\mathrm{PV}}$ reflecting rooftop configuration, shading conditions, infrastructure shadows, weather conditions, and surrounding visual features. The automated extraction stage employed a pretrained vision–language model operating under a predefined structured output format to ensure consistency and compatibility with downstream data processing (Fig.~\ref{fig:Automatic_Labeling_Prompt}).

The application of vision–language models for structured semantic extraction has become increasingly common in remote sensing analysis. Such tools enable scalable annotation of rooftop attributes that would otherwise require extensive manual inspection. Prior studies including PVAL~\cite{guo2025solar}, MiniGPT-4~\cite{zhu2023minigpt}, and LLaVA~\cite{liu2023visual} demonstrate the effectiveness of instruction-guided visual interpretation for structured data construction. In this implementation, automated annotations serve as a preliminary labeling stage. To ensure reliability suitable for planning applications, all automated annotations were subjected to expert verification by four independent domain-aware reviewers. Reviewers examined extracted labels, corrected misclassifications, resolved inconsistencies, and adjudicated ambiguous cases arising from occlusions or rooftop complexity. Disagreements were resolved through reviewer agreement after further examination. This validation framework combines automated scalability with expert oversight, resulting in a dependable rooftop reference repository for operational PV assessment.

\begin{figure}[h]
\centering
\includegraphics[width=1\columnwidth]{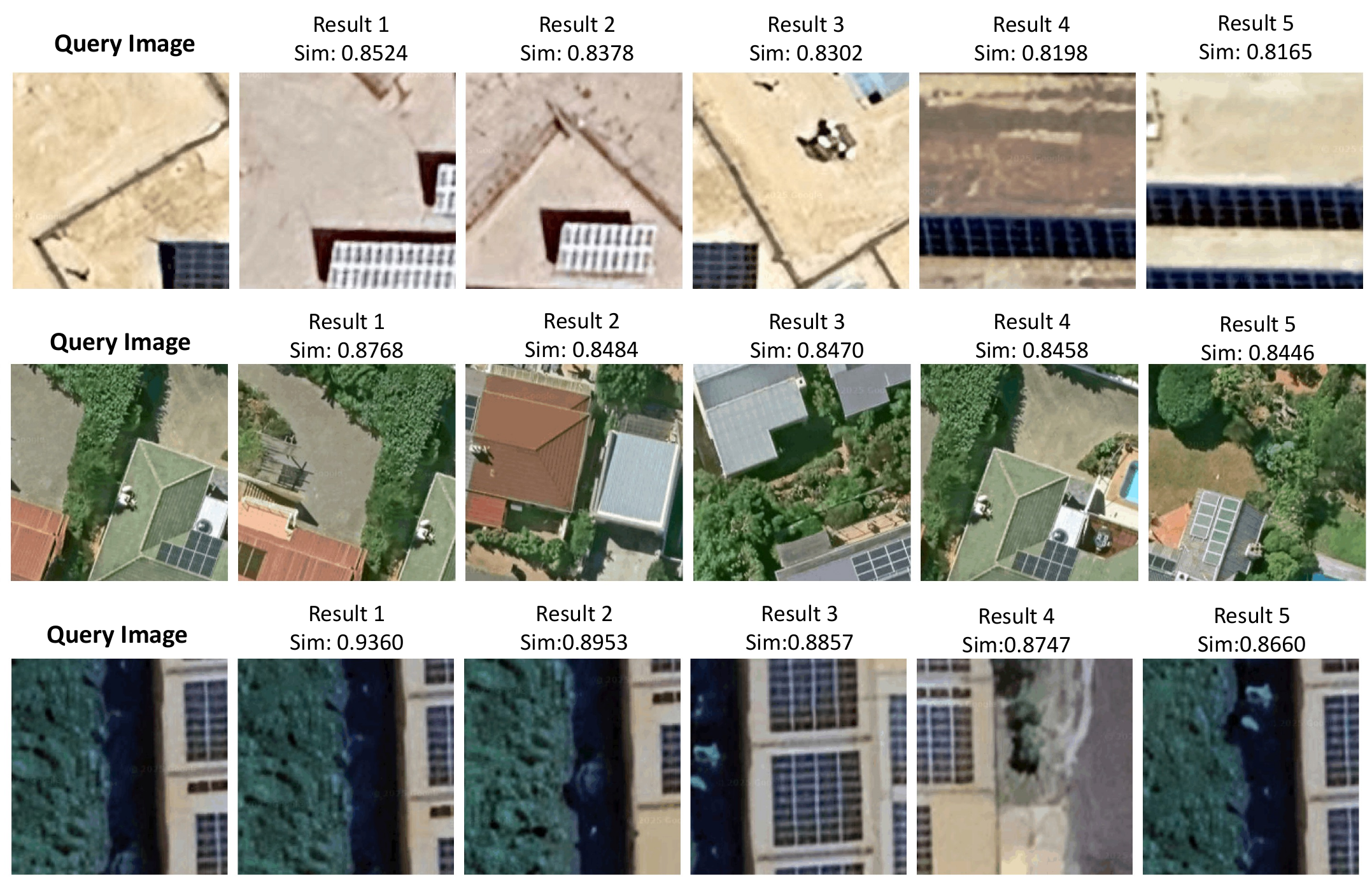}
\caption{Representative examples of similarity-based reference selection for rooftop PV assessment. For each input satellite image, the five most similar reference rooftop cases are identified using normalized numerical similarity measures. The similarity values reflect strong correspondence in rooftop geometry, panel appearance, and surrounding environmental conditions.}
\label{fig:Query_Similarity}
\end{figure}

All validated satellite images, together with their structured PV descriptors and associated rooftop attributes, were transformed into numerical similarity representations using a vision–language model. These representations were organized within a searchable reference database to enable efficient similarity-based selection of comparable rooftop cases during PV estimation (see Fig.~\ref{fig:Query_Similarity}). The developed rooftop reference database is designed to support additional expansion. Additional cities, rooftop configurations, or emerging PV technologies may be incorporated by adding newly validated satellite images without altering the PV estimation framework $f_{\mathrm{PV}}(\cdot)$. This separation between data updates and the estimation procedure enables scalable application across multiple distribution service territories. To promote transparency and reproducibility, the implementation scripts and curated dataset will be made publicly available upon manuscript acceptance.

\subsection{Similarity-Based Rooftop Reference Retrieval}
\label{subsec:retrieval}

Given an input satellite image $X_q$, the first stage of the procedure computes a numerical similarity representation using a pretrained vision–language model (CLIP ViT-B/32). The model maps the image to a 512-dimensional numerical vector
\begin{equation}
\mathbf{h}_q = \phi(X_q),
\end{equation}
where $\phi(\cdot)$ denotes the image-to-vector transformation. To ensure consistent similarity comparison across regions and deployment conditions, all vectors are normalized using the Euclidean norm,
\begin{equation}
\tilde{\mathbf{h}}_q = \frac{\mathbf{h}_q}{\|\mathbf{h}_q\|_2}.
\end{equation}

The same normalization procedure is applied to all reference satellite images $X_j^{r}$ in the database, producing normalized vectors $\tilde{\mathbf{h}}_j$. Similarity-based reference selection is then performed using a numerical distance metric. For a given input image $X_q$, the $K$ most similar reference images are identified by minimizing the Euclidean distance between normalized vectors,
\begin{equation}
\mathcal{R}(X_q)
=
\operatorname*{arg\,topK}_{X_j^{r} \in \mathcal{K}}
\left(
-\left\| \tilde{\mathbf{h}}_q - \tilde{\mathbf{h}}_j \right\|_2
\right).
\end{equation}

Because the vectors are normalized, minimizing Euclidean distance is equivalent to maximizing cosine similarity, ensuring consistent similarity comparison across images. This similarity-based matching allows the method to reference rooftop cases with comparable geometric structures, shading patterns, and surrounding environmental conditions. Consequently, the PV estimation results are less sensitive to regional differences in building practices, climate conditions, and variations in image quality.

For clarity in reporting, the similarity value is defined as
\begin{equation}
\mathrm{sim}(X_q, X_j^{r}) = \frac{1}{1 + d(X_q, X_j^{r})},
\end{equation}
where
\begin{equation}
d(X_q, X_j^{r}) = \left\| \tilde{\mathbf{h}}_q - \tilde{\mathbf{h}}_j \right\|_2.
\end{equation}

\subsection{Solar-RAG Reference-Assisted PV Estimation}
\label{subsec:rag_inference}

\begin{figure}[t]
\centering
\includegraphics[width=\columnwidth]{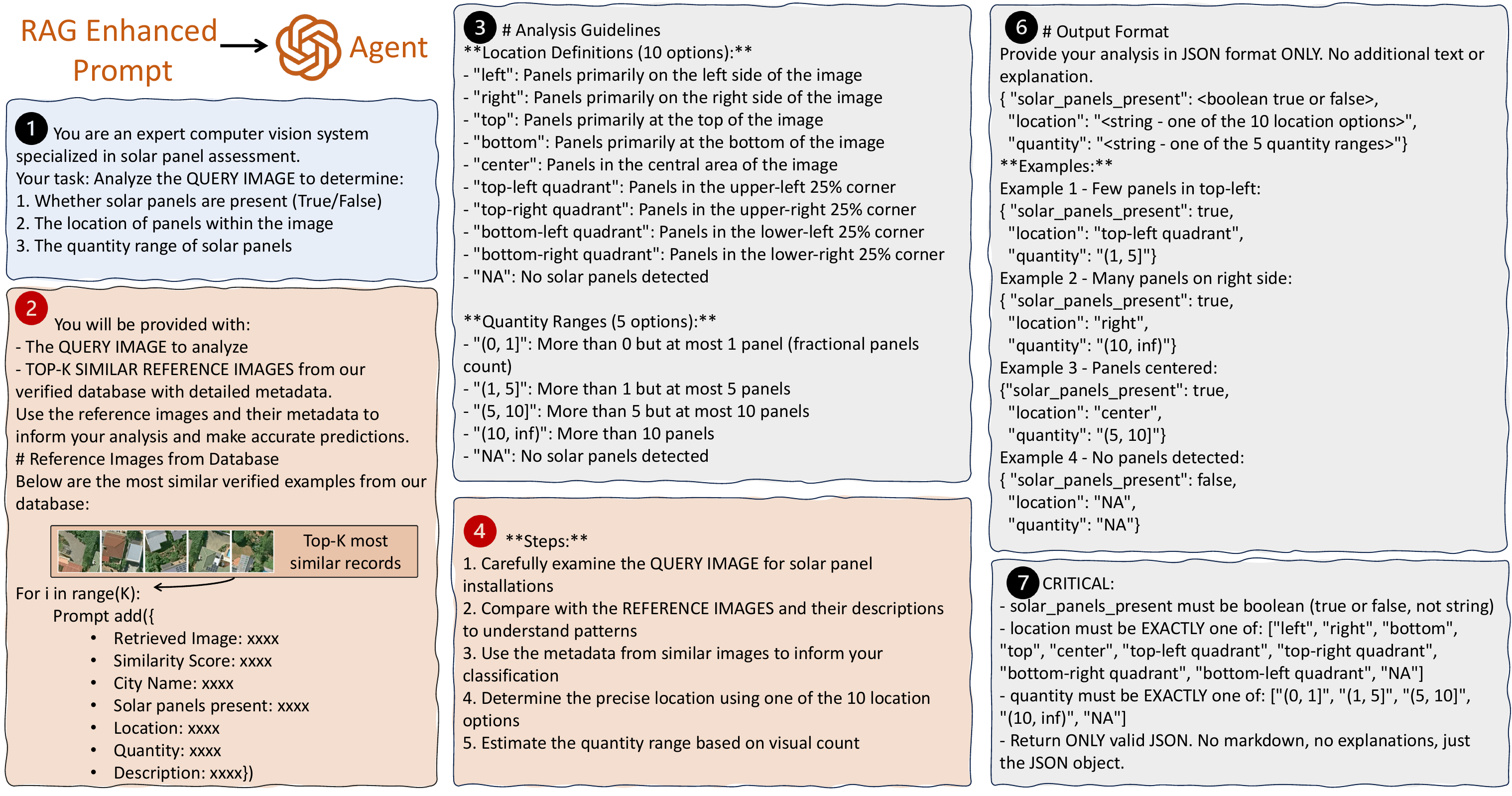}
\caption{Retrieval-augmented prompt structure used for PV assessment.}
\label{fig:RAG_Enhanced_Prompt}
\end{figure}

The selected reference cases $\mathcal{R}(X_q)$ provide supporting examples for PV estimation. Each reference image includes validated PV descriptors ($p_j^{\mathrm{PV}}$, $n_j^{\mathrm{PV}}$, $\ell_j^{\mathrm{PV}}$, $e_j^{\mathrm{PV}}$) that describe rooftop geometry and PV installation characteristics. During estimation, the input image $X_q$, together with the selected reference images and their corresponding PV descriptors $\mathbf{z}_j^{\mathrm{PV}}$, are provided to the PV estimation model through a predefined input structure (Fig.~\ref{fig:RAG_Enhanced_Prompt}). The model performs a comparative assessment between the input rooftop and the reference rooftops to determine the structured PV descriptor vector
\begin{equation}
\hat{\mathbf{z}}^{\mathrm{PV}} =
\left[
\hat{p}^{\mathrm{PV}},
\hat{n}^{\mathrm{PV}},
\hat{\ell}^{\mathrm{PV}},
\hat{e}^{\mathrm{PV}}
\right].
\end{equation}

By using visually and geographically comparable rooftop cases as references, the proposed approach reduces sensitivity to regional differences in construction practices, climate conditions, and variations in image quality. The structured PV descriptors produced by the framework enable feeder-level aggregation of distributed PV installations, supporting distribution system modeling, hosting capacity assessment, and distributed energy resource integration studies. This reference-based estimation framework minimizes the need for repeated model updates when applied to new service territories while preserving structured outputs compatible with techno-economic analysis in distribution system planning.

\section{Experimental Setup}
\label{sec:Experiments}

This section describes the experimental design used to evaluate the proposed Solar-RAG framework for satellite-based PV inventory estimation. The evaluation focuses on three aspects that are critical for distribution-system applications: geographic robustness across diverse rooftop environments, accuracy of structured PV descriptor estimation, and the contribution of retrieval-guided contextual reasoning relative to standalone vision--language and conventional computer vision baselines.

\paragraph{\textbf{Dataset construction and evaluation tasks}}
The experiments are conducted on a globally distributed satellite image dataset constructed from open and publicly accessible sources. Each satellite image corresponds to a standardized overhead view of a residential or urban area and is annotated with structured PV attributes. Specifically, each image is labeled for solar panel presence (binary), panel location within the image (region- or quadrant-based categories), and panel quantity expressed as an interval from the set $(0,1]$, $(1,5]$, $(5,10]$, $(10,+\infty)$, or ``NA'' when no PV is present. These annotations provide a standardized and interpretable description of rooftop PV characteristics across different geographic regions. Together, these tasks provide the structured information required for rooftop PV inventory construction, feeder-level aggregation, and estimation of distributed generation penetration.

Table~\ref{tab:dataset_split_summary} summarizes the datasets used for evaluation and reference matching. For each region, a total of 480 satellite images are randomly partitioned into two disjoint subsets of equal size: 240 images are used exclusively for performance evaluation, and the remaining 240 images are used to construct the reference database. This protocol is applied consistently across regions in North America, Europe, Asia, Africa, Oceania, and South America, enabling controlled cross-regional evaluation while preventing any overlap between evaluation images and reference examples.

\begin{table}[h]
\centering
\caption{Regional dataset split for evaluation and retrieval.}
\renewcommand{\arraystretch}{1.1}
\setlength{\tabcolsep}{6pt}
\begin{adjustbox}{width=\columnwidth,center}
\begin{tabular}{llccc}
\hline
\hline
\textbf{Continent} & \textbf{Region / City} & \textbf{Evaluation} & \textbf{RAG Reference} & \textbf{Total} \\
\textbf{Asia (Middle East)} & Kuwait City, Kuwait & 240 & 240 & 480 \\
\textbf{Oceania} & Sydney, Australia & 240 & 240 & 480 \\
\textbf{Africa} & Cape Town, South Africa & 240 & 240 & 480 \\
\textbf{Europe} & Oxford, UK & 240 & 240 & 480 \\
\textbf{South America} & S\~{a}o Paulo, Brazil & 240 & 240 & 480 \\
\textbf{Asia (East Asia)} & Shanghai, China & 240 & 240 & 480 \\
\hline
\multirow{6}{*}{\textbf{North America}}
 & Tempe, AZ        & 240 & 240 & 480 \\
 & Tacoma, WA       & 240 & 240 & 480 \\
 & Seattle, WA      & 240 & 240 & 480 \\
 & Orlando, FL      & 240 & 240 & 480 \\
 & Osage Beach, MO  & 240 & 240 & 480 \\
 & Harlem, NY       & 240 & 240 & 480 \\
\hline
\end{tabular}
\end{adjustbox}
\label{tab:dataset_split_summary}
\end{table}

\paragraph{\textbf{Model configuration and retrieval setting}}

The proposed framework employs a multimodal vision--language model for structured PV descriptor estimation, while similarity-assisted inference is implemented through retrieval of comparable rooftop reference cases~\cite{openai_api}. For similarity-assisted estimation, each query satellite image is encoded using a CLIP ViT-B/32 image encoder~\cite{radford2021learning}, producing a 512-dimensional L2-normalized numerical representation. These representations are compared against a FAISS-indexed~\cite{douze2025faiss} reference database consisting of geographically distributed rooftop satellite images with validated PV descriptors.
During estimation, the top-$K$ most similar reference images (with $K=3$ unless otherwise specified) are selected based on Euclidean distance in the normalized representation space. Their similarity scores, geographic identifiers, and structured PV descriptors are incorporated into the estimation prompt to support comparative assessment of rooftop characteristics.
For benchmarking purposes, all experiments were executed on the \textit{Sol} supercomputing cluster~\cite{jennewein2023sol} at Arizona State University. Each computational node is equipped with NVIDIA A100 GPUs (80~GB memory, CUDA~12.7), AMD EPYC 7413 CPUs, and 256~GB of system memory. Baseline computer vision models, including U-Net~\cite{bouaziz2024high}, ResNet-152~\cite{he2016deep}, Inception-v3~\cite{szegedy2016rethinking}, VGG-19~\cite{simonyan2014very}, and ViT-Base-16~\cite{dosovitskiy2020image}, were implemented under identical data preprocessing, augmentation, and batch-size settings to ensure a consistent and fair performance comparison with the proposed vision-language-based PV assessment approach.

We evaluate three implementation variants: 
(1) \textbf{Similarity-assisted GPT-4o (proposed)}, which performs PV estimation using reference supported reasoning based on retrieved rooftop cases; 
(2) \textbf{GPT-4o}~\cite{achiam2023gpt}, which processes only the query satellite image and task description without reference support; and 
(3) \textbf{GPT-5.2}~\cite{openai_gpt5_2}, evaluated under the same standalone configuration.  
All models are instructed to produce structured outputs in a fixed JSON format containing PV presence, spatial location, quantity interval, and a concise explanatory description. The similarity-based reference selection is applied strictly during the estimation stage and does not modify model parameters. Consequently, observed performance differences reflect the impact of reference-supported estimation rather than additional training or parameter adjustment. All experiments are implemented in Python~3.9 using FAISS for similarity search. Identical image preprocessing procedures, prompt structures, and output formats are applied across all model variants to ensure a consistent and equitable comparison.

\paragraph{\textbf{Evaluation protocol}}
Model performance is assessed on three structured PV assessment tasks:
(i) PV presence identification,
(ii) PV spatial localization, and
(iii) panel quantity interval estimation. 
For all tasks, predictions are evaluated using an exact-match criterion: a prediction is considered correct only if it exactly matches the corresponding ground-truth label. For quantity estimation, correctness is defined as selecting the appropriate predefined interval.

PV presence identification is formulated as a binary classification problem. In addition to overall accuracy, precision, recall, and F1-score are reported to quantify the balance between false positive and false negative outcomes. Let $\mathrm{TP}$, $\mathrm{FP}$, and $\mathrm{FN}$ denote the numbers of true positives, false positives, and false negatives, respectively. The metrics are defined as
\begin{equation}
\mathrm{Precision} = \frac{\mathrm{TP}}{\mathrm{TP} + \mathrm{FP}},
\end{equation}
\begin{equation}
\mathrm{Recall} = \frac{\mathrm{TP}}{\mathrm{TP} + \mathrm{FN}},
\end{equation}
\begin{equation}
\mathrm{F1} = \frac{2 \cdot \mathrm{Precision} \cdot \mathrm{Recall}}
{\mathrm{Precision} + \mathrm{Recall}}.
\end{equation}

These metrics provide a comprehensive performance assessment, particularly for cases involving low image clarity, partial shading, or visually complex rooftop configurations.
PV spatial localization and panel quantity interval estimation are evaluated using task-specific accuracy, consistent with their discrete label definitions. All evaluations are performed exclusively on the held-out subsets summarized in Table~\ref{tab:dataset_split_summary}. To avoid information leakage, all reference images used for similarity matching are drawn solely from the disjoint reference subsets. Results are reported both at the regional level and in aggregate across all regions to assess performance consistency under varying geographic and architectural conditions. In addition to quantitative metrics, qualitative case analyses are conducted on representative examples to examine model behavior under challenging scenarios, including partial occlusion, rooftop clutter, and ambiguous visual patterns.

\section{Results}
\label{sec:results}

\subsection{Qualitative Case Studies: Interpretation of Similarity-Assisted Estimation}
\label{subsec:qualitative_cases}

\begin{figure}[h]
\centering
\includegraphics[width=1\columnwidth]{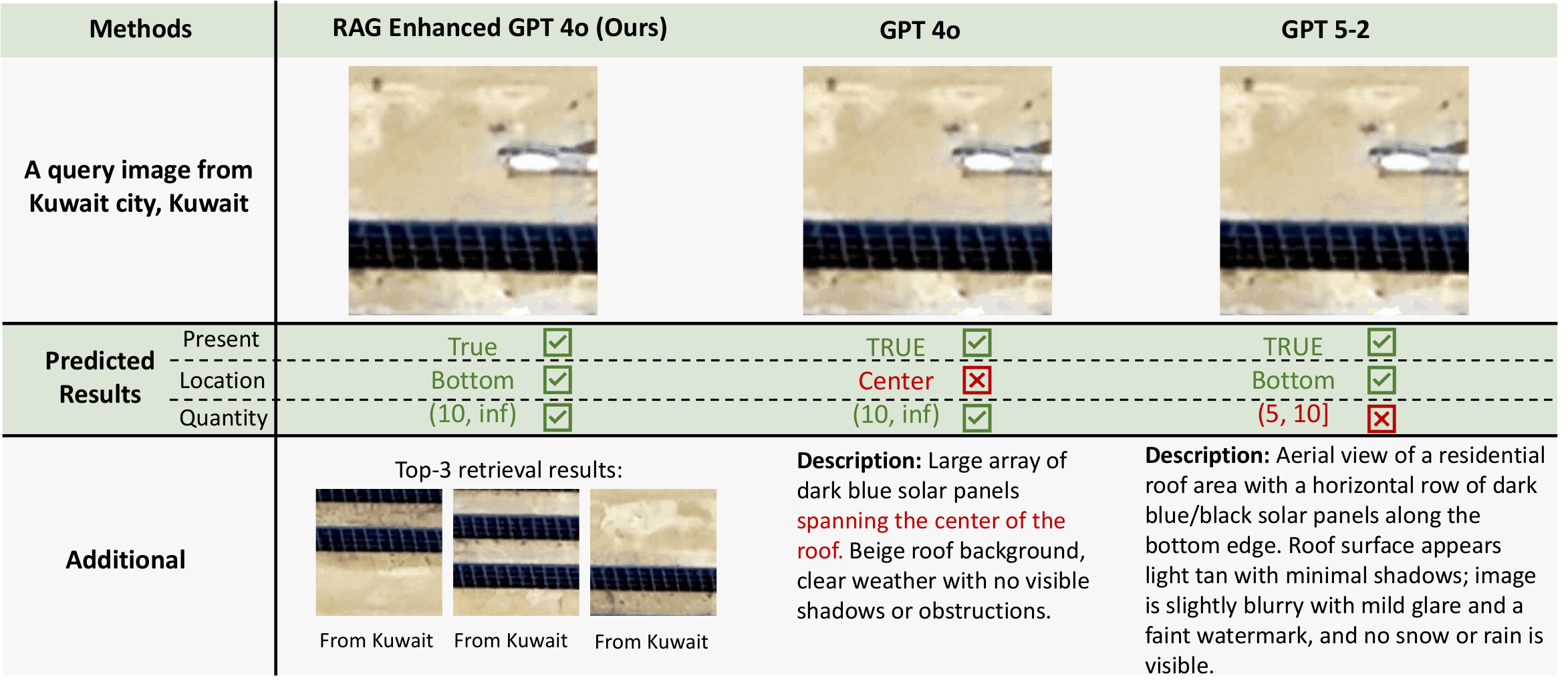}
\caption{Qualitative analysis of a satellite image from Kuwait City, Kuwait, where PV panels are installed along the lower portion of the rooftop. All models correctly detect PV presence. However, GPT-4o and GPT-5.2 exhibit inaccuracies in spatial localization and quantity estimation. The similarity-assisted GPT-4o, supported by regionally comparable reference cases, correctly identifies the array location at the lower roof boundary and provides a consistent quantity interval estimate. In contrast, the standalone models misidentify the panel position or underestimate the number of installed modules.}
\label{fig:Case_analysis_Kuwait}
\end{figure}

\begin{figure}[h]
\centering
\includegraphics[width=1\columnwidth]{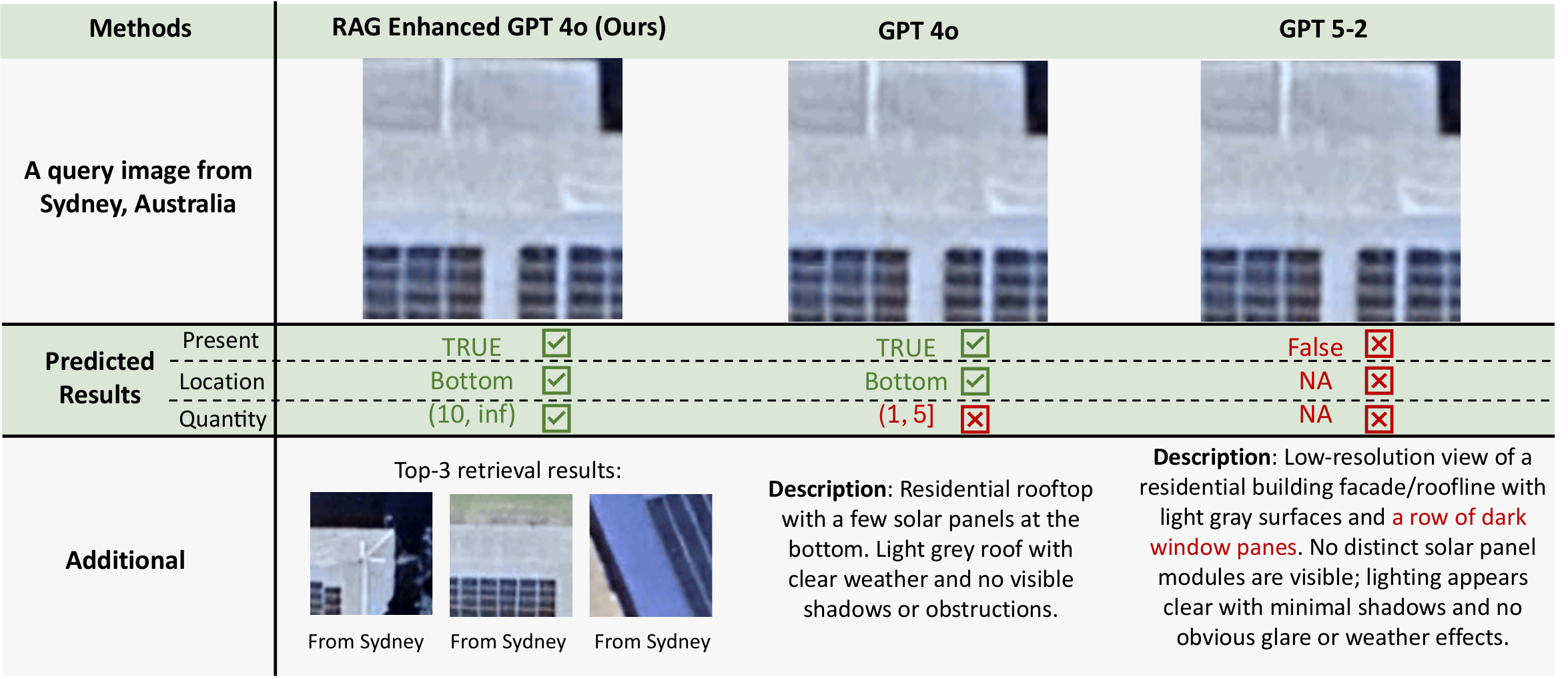}
\caption{Qualitative comparison for a rooftop image from Sydney, Australia. The similarity-assisted GPT-4o correctly identifies PV presence, bottom-region localization, and quantity interval $(10, \infty)$ using the top-3 retrieved regional reference cases. GPT-4o without similarity support underestimates the panel quantity. GPT-5.2 produces a false negative by misclassifying PV modules as facade or window-like structural elements.}
\label{fig:Case_analysis_Sydney}
\end{figure}

\begin{figure}[h]
\centering
\includegraphics[width=1\columnwidth]{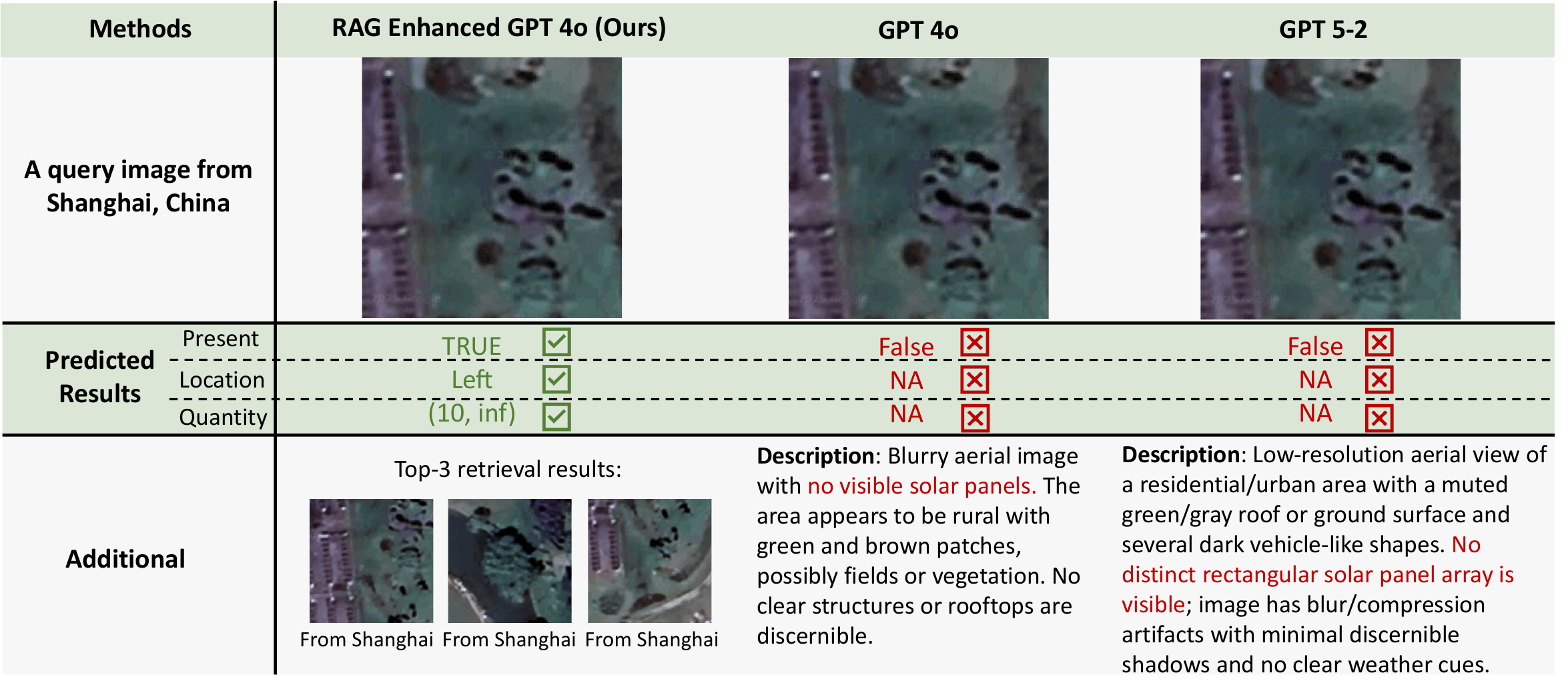}
\caption{Qualitative analysis of a low-resolution satellite image from Shanghai, China. The similarity-assisted GPT-4o predicts PV presence on the left region with quantity interval $(10,\infty)$, guided by retrieved regional cases exhibiting faint but consistent PV patterns. In contrast, GPT-4o and GPT-5.2 both produce negative predictions and state that no visible solar panels are detected, attributing the scene to image blur, vegetation, or vehicle-like structures. As a result, the standalone models fail to recognize subtle PV indicators under reduced image clarity.}
\label{fig:Case_analysis_Shanghai}
\end{figure}

\begin{figure}[h]
\centering
\includegraphics[width=1\columnwidth]{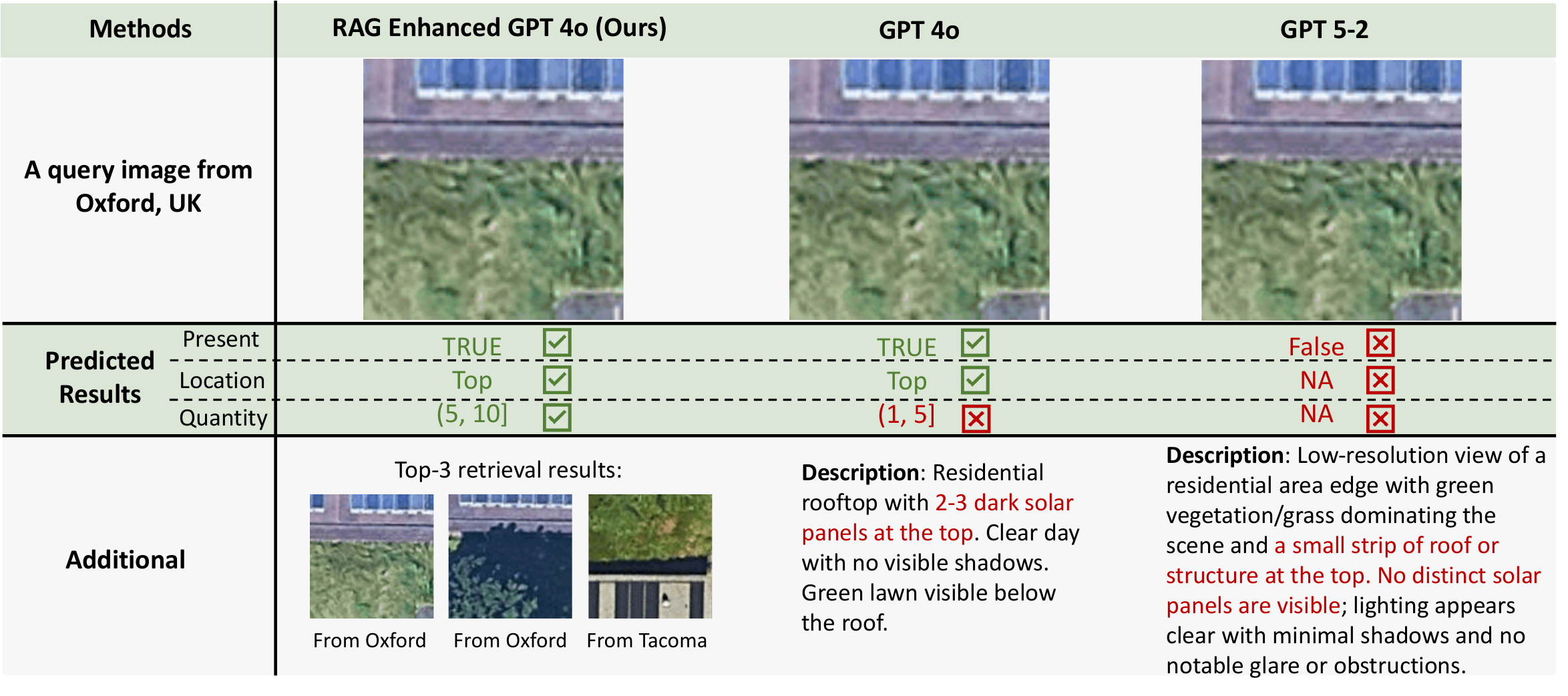}
\caption{Qualitative analysis of a rooftop image from Oxford, UK. The similarity-assisted GPT-4o correctly identifies PV presence, top-region localization, and quantity interval $(5,10)$ using reference rooftops with comparable spatial layouts. GPT-4o without similarity support detects PV presence and location but underestimates the quantity interval as $(1,5)$, consistent with its description emphasizing ``2--3 dark solar panels at the top.'' GPT-5.2 produces a false negative by interpreting the upper region as non-PV roof structure rather than installed modules.}
\label{fig:Case_analysis_Oxford}
\end{figure}

\begin{figure}[h]
\centering
\includegraphics[width=1\columnwidth]{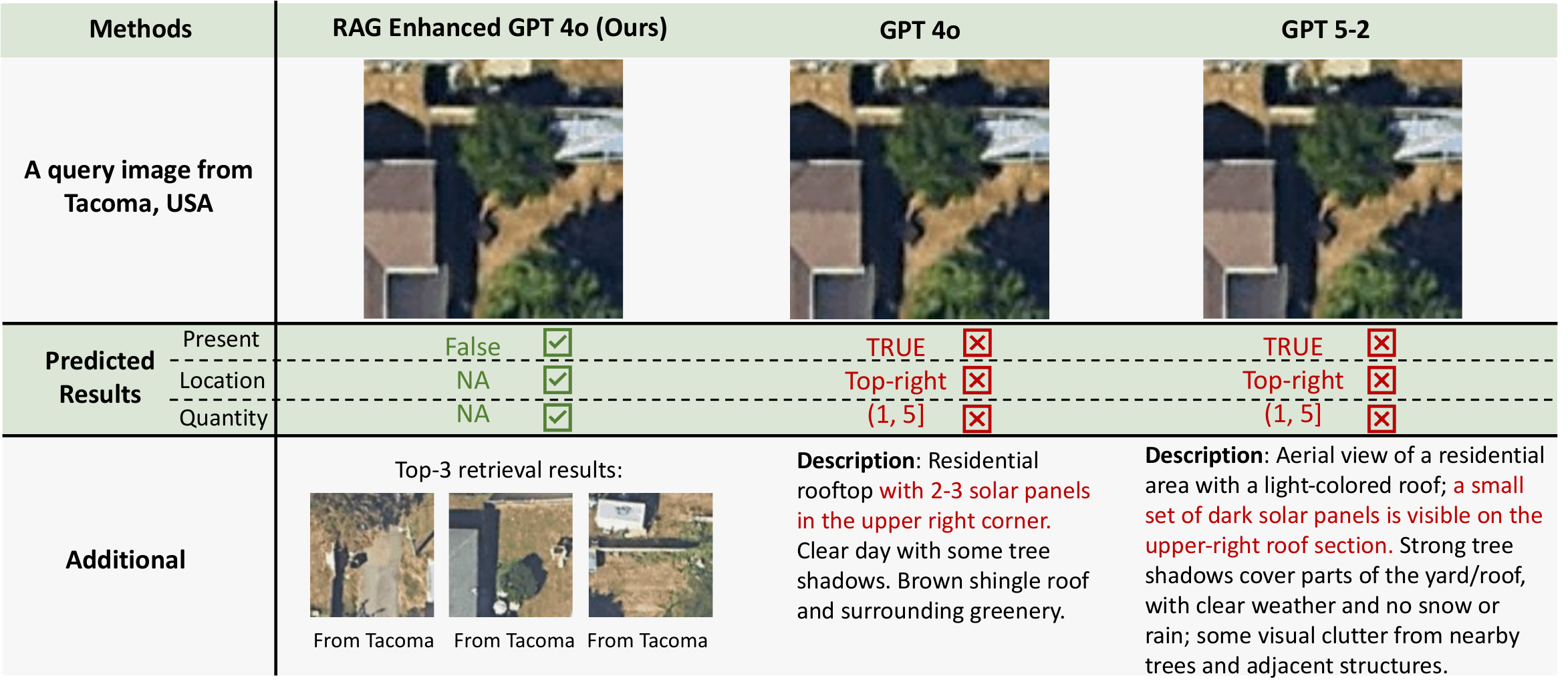}
\caption{Qualitative analysis of a residential rooftop image from Tacoma, USA, where no PV panels are installed. GPT-4o and GPT-5.2 both produce false positive predictions and describe apparent PV-like features (e.g., ``2-3 dark solar panels in the upper-right corner''), indicating sensitivity to dark or rectangular roof textures caused by shading or structural elements. In contrast, the similarity-assisted GPT-4o does not generate a panel-oriented interpretation. Comparison with retrieved regionally similar non-PV rooftops provides consistent negative visual evidence, leading to a correct no-PV classification.}
\label{fig:Case_analysis_Tacoma}
\end{figure}

To better understand how similarity-based reference selection influences estimation behavior beyond aggregate performance metrics, we present five representative qualitative case studies drawn from different geographic regions and imaging conditions: Kuwait City, Sydney, Shanghai, Oxford, and Tacoma. Each case compares the proposed similarity-assisted GPT-4o with its standalone counterparts (GPT-4o and GPT-5.2), examining both structured outputs (PV presence, spatial location, and quantity interval) and the associated descriptive explanations. Analysis of these explanations provides insight into the visual features emphasized by each model and highlights consistent patterns in performance improvements and failure modes.

The Kuwait City case (Fig.~\ref{fig:Case_analysis_Kuwait}) illustrates improved spatial and quantitative estimation when PV installations are clearly present. In this example, a large horizontal array is installed along the lower boundary of the rooftop. All models correctly identify PV presence. However, GPT-4o incorrectly localizes the array to the center of the roof, and GPT-5.2 underestimates the quantity interval as $(5,10]$. In contrast, the similarity-assisted GPT-4o retrieves comparable rooftop configurations from the same geographic region and correctly localizes the array to the lower portion of the roof while assigning the appropriate $(10,\infty)$ quantity interval. This case demonstrates that regional reference information improves not only detection accuracy but also spatial and numerical consistency.

In the Sydney case (Fig.~\ref{fig:Case_analysis_Sydney}), the rooftop contains an extended PV array distributed across the lower section. The similarity-assisted GPT-4o correctly predicts PV presence, bottom-region localization, and a large quantity interval. GPT-4o without reference support underestimates the quantity by focusing on the most visually prominent cluster of panels, while GPT-5.2 produces a false negative by misclassifying the panels as structural facade elements. Incorporation of regionally comparable rooftop examples enables the similarity-assisted model to account for the full spatial extent of the installation rather than relying solely on the most salient visual region.

The Shanghai case (Fig.~\ref{fig:Case_analysis_Shanghai}) demonstrates estimation behavior under reduced image clarity. The satellite image exhibits low resolution and blur, with panels appearing as faint rectangular features. The similarity-assisted GPT-4o predicts PV presence on the left region and assigns a large quantity interval. In contrast, both GPT-4o and GPT-5.2 produce negative predictions, emphasizing blur, compression artifacts, or non-PV objects in their explanations. In this scenario, standalone models appear to rely heavily on overall image clarity when forming decisions. By retrieving regionally comparable examples containing subtle but valid PV patterns, the similarity-assisted model interprets faint rectangular structures as panels rather than background artifacts.

The Oxford case (Fig.~\ref{fig:Case_analysis_Oxford}) involves a rooftop with a moderate number of panels installed near the roof edge. The similarity-assisted GPT-4o correctly predicts PV presence, top-region localization, and a quantity interval of $(5,10)$. GPT-4o without reference support detects presence and location but underestimates the quantity interval as $(1,5)$, consistent with its description highlighting only a subset of visible panels. GPT-5.2 produces a false negative by interpreting the upper region as roof structure or surrounding vegetation. Retrieved examples showing similar roof-edge installations enable the similarity-assisted model to account for additional panels that may not dominate the visual scene.

The Tacoma case (Fig.~\ref{fig:Case_analysis_Tacoma}) illustrates false-positive behavior in the absence of PV installations. The ground truth contains no panels; however, both GPT-4o and GPT-5.2 predict PV presence, localize panels to the upper-right region, and assign non-zero quantity intervals. Their descriptions indicate interpretation of dark rectangular textures, likely caused by shading or roof geometry, as PV modules. In contrast, the similarity-assisted GPT-4o produces a correct negative prediction and does not generate panel-related interpretation. Comparison with regionally similar non-PV rooftops allows the model to attribute these dark features to structural elements rather than installed PV systems.

Across all five cases, a consistent pattern is observed. Standalone models rely predominantly on local texture cues, visually prominent regions, or overall image clarity when forming predictions, which can result in underestimation, missed detection, or false positives depending on scene characteristics. The descriptive explanations reveal which visual factors dominate these decisions. Incorporating similarity-based regional references provides additional contextual evidence, enabling reinterpretation of ambiguous patterns and improving consistency in spatial localization and quantity estimation. These case studies support the quantitative findings by illustrating how reference-assisted estimation enhances robustness under diverse rooftop and imaging conditions.

\subsection{Quantitative Results Analysis}
\label{subsec:Results}

\begin{table}[t]
\centering
\caption{City-level photovoltaic assessment accuracy (\%). Results are reported for
photovoltaic presence detection (\emph{Present}), panel quantity estimation
(\emph{Quantity}), and spatial localization (\emph{Location}).}
\label{tab:results_citywise}
\adjustbox{max width=\columnwidth}{
\begin{tabular}{lccc|ccc|ccc}
\hline
\multirow{2}{*}{City} &
\multicolumn{3}{c}{RAG-Enhanced (Ours)} &
\multicolumn{3}{c}{GPT-4o} &
\multicolumn{3}{c}{GPT-5.2} \\
\cline{2-10}
 & Presence & Quantity & Location
 & Presence & Quantity & Location
 & Presence & Quantity & Location \\
\hline
\rowcolor{blue!15}
\textbf{Overall}
 & \textbf{95.4 $\pm$ 0.7} & \textbf{86.2 $\pm$ 0.7} & \textbf{80.2 $\pm$ 1.6}
 & 93.6 $\pm$ 0.9 & 83.4 $\pm$ 2.2 & 79.5 $\pm$ 2.1
 & 85.9 $\pm$ 1.4 & 67.7 $\pm$ 2.2 & 71.0 $\pm$ 1.7 \\
\hline
Kuwait City, Kuwait
 & 100.0 $\pm$ 0.8 & 95.2 $\pm$ 0.3 & 88.1 $\pm$ 1.3
 & 100.0 $\pm$ 0.7 & 90.5 $\pm$ 1.1 & 78.6 $\pm$ 1.5
 & 100.0 $\pm$ 0.7 & 54.8 $\pm$ 2.2 & 90.5 $\pm$ 1.0 \\
São Paulo, Brazil
 & 92.2 $\pm$ 1.2 & 77.9 $\pm$ 2.3 & 64.9 $\pm$ 3.6
 & 92.2 $\pm$ 0.9 & 72.7 $\pm$ 1.2 & 59.7 $\pm$ 2.6
 & 83.1 $\pm$ 1.5 & 55.8 $\pm$ 2.4 & 64.9 $\pm$ 3.9 \\
Sydney, Australia
 & 100.0 $\pm$ 0.8 & 73.9 $\pm$ 1.6 & 78.3 $\pm$ 2.1
 & 100.0 $\pm$ 0.5 & 69.6 $\pm$ 3.8 & 69.6 $\pm$ 2.9
 & 87.0 $\pm$ 0.9 & 52.2 $\pm$ 2.5 & 69.6 $\pm$ 3.1 \\
Cape Town, South Africa
 & 98.6 $\pm$ 0.5 & 80.6 $\pm$ 2.0 & 59.7 $\pm$ 3.8
 & 97.2 $\pm$ 0.5 & 72.2 $\pm$ 1.5 & 62.5 $\pm$ 4.0
 & 91.7 $\pm$ 1.1 & 65.3 $\pm$ 2.2 & 58.3 $\pm$ 2.1 \\
Oxford, UK
 & 92.8 $\pm$ 1.1 & 75.4 $\pm$ 2.4 & 75.4 $\pm$ 2.3
 & 91.3 $\pm$ 0.9 & 68.1 $\pm$ 3.0 & 69.6 $\pm$ 2.4
 & 85.5 $\pm$ 1.4 & 53.6 $\pm$ 3.7 & 53.6 $\pm$ 2.6 \\
Shanghai, China
 & 93.9 $\pm$ 0.8 & 90.5 $\pm$ 1.2 & 72.8 $\pm$ 2.2
 & 91.2 $\pm$ 1.0 & 87.1 $\pm$ 0.8 & 72.1 $\pm$ 1.7
 & 59.2 $\pm$ 4.0 & 50.3 $\pm$ 3.1 & 48.3 $\pm$ 3.9 \\
\cline{1-1}
Tempe, AZ, USA
 & 98.3 $\pm$ 0.3 & 95.7 $\pm$ 0.4 & 95.3 $\pm$ 0.6
 & 99.1 $\pm$ 0.3 & 93.5 $\pm$ 0.9 & 97.0 $\pm$ 0.7
 & 97.4 $\pm$ 0.6 & 83.6 $\pm$ 1.5 & 87.9 $\pm$ 1.2 \\
Orlando, FL, USA
 & 97.1 $\pm$ 0.8 & 85.3 $\pm$ 0.7 & 94.1 $\pm$ 1.3
 & 100.0 $\pm$ 0.7 & 100.0 $\pm$ 0.5 & 100.0 $\pm$ 0.4
 & 91.2 $\pm$ 1.5 & 76.5 $\pm$ 1.6 & 82.4 $\pm$ 1.3 \\
Seattle, WA, USA
 & 91.9 $\pm$ 0.8 & 80.6 $\pm$ 2.3 & 79.0 $\pm$ 2.0
 & 85.5 $\pm$ 1.3 & 77.4 $\pm$ 2.1 & 77.4 $\pm$ 1.9
 & 85.5 $\pm$ 1.5 & 75.8 $\pm$ 1.7 & 77.4 $\pm$ 1.9 \\
Osage Beach, MO, USA
 & 76.2 $\pm$ 1.5 & 57.1 $\pm$ 2.2 & 61.9 $\pm$ 2.6
 & 66.7 $\pm$ 3.3 & 66.7 $\pm$ 2.7 & 66.7 $\pm$ 2.7
 & 76.2 $\pm$ 1.5 & 66.7 $\pm$ 2.5 & 61.9 $\pm$ 3.9 \\
Harlem, NY, USA
 & 100.0 $\pm$ 0.7 & 85.3 $\pm$ 1.2 & 79.4 $\pm$ 1.4
 & 91.2 $\pm$ 1.3 & 91.2 $\pm$ 0.8 & 91.2 $\pm$ 1.0
 & 91.2 $\pm$ 1.5 & 73.5 $\pm$ 2.0 & 73.5 $\pm$ 1.9 \\
Tacoma, WA, USA
 & 92.0 $\pm$ 1.4 & 84.0 $\pm$ 2.0 & 86.0 $\pm$ 1.4
 & 86.0 $\pm$ 1.2 & 78.0 $\pm$ 2.1 & 80.0 $\pm$ 1.3
 & 92.0 $\pm$ 0.9 & 84.0 $\pm$ 1.6 & 82.0 $\pm$ 1.3 \\
\hline
\end{tabular}
}
\end{table}

Table~\ref{tab:results_citywise} summarizes city-level performance for the structured PV inventory estimation tasks, including PV presence identification, panel quantity interval estimation, and spatial localization. These results characterize how reliably the proposed framework can recover rooftop PV descriptors across diverse geographic environments. To ensure metric reliability and reproducibility, three independent experimental runs were conducted for each configuration, and the mean accuracy with the corresponding standard deviation ($\pm \sigma$) is reported.

The similarity-assisted model achieves the highest overall accuracy across all three tasks, with $95.4\% \pm 0.7\%$ for presence identification, $86.2\% \pm 0.7\%$ for quantity estimation, and $80.2\% \pm 1.6\%$ for spatial localization. In comparison, GPT-4o without reference support attains $93.6\% \pm 0.9\%$, $83.4\% \pm 2.2\%$, and $79.5\% \pm 2.1\%$, respectively. GPT-5.2 exhibits substantially lower performance, particularly for quantity estimation and spatial localization. These aggregate results indicate that incorporation of regionally comparable reference cases improves both numerical consistency and spatial reliability of PV assessment across geographically diverse conditions.

Presence identification accuracy is high across most cities for all models, indicating that binary PV detection is generally less demanding than the other tasks. Several regions demonstrate near-perfect presence accuracy, including Sydney, Kuwait City, and Harlem. Even under these high-accuracy conditions, the similarity-assisted model maintains a consistent advantage in the overall aggregate, suggesting improved detection consistency when applied at scale. GPT-5.2 exhibits greater variability across cities, achieving strong performance in some regions but showing marked degradation in others. The most notable example is Shanghai, where presence accuracy decreases to 59.2\%, compared to over 90\% for the other two models.

\textit{Panel quantity interval estimation} reveals a clearer performance separation among methods. The similarity-assisted model consistently achieves the highest quantity accuracy across most cities and delivers the strongest aggregate performance. Improvements are particularly evident in dense urban environments such as Shanghai and São Paulo, where rooftop layouts are complex and background structures are visually dense. Under these conditions, similarity-based reference information appears to support more accurate interpretation of installation size. GPT-4o performs competitively but shows reduced accuracy in several regions, while GPT-5.2 demonstrates significant performance degradation, indicating difficulty in maintaining consistent numerical estimation across different urban and architectural settings.

\textit{PV spatial localization} remains the most challenging task, with lower absolute accuracies and greater variability across cities for all models. Nevertheless, the similarity-assisted approach achieves the highest overall localization accuracy and demonstrates consistent improvements in cities such as Tempe, Orlando, Tacoma, and Kuwait City. GPT-4o attains comparable but slightly lower aggregate performance, whereas GPT-5.2 again shows reduced reliability, particularly in visually dense urban environments. These findings suggest that referencing visually comparable rooftop configurations provides useful spatial guidance that enhances localization accuracy.

Overall, the experimental results demonstrate consistent performance improvements across all evaluated tasks when similarity-based regional reference information is incorporated. The largest gains are observed in panel quantity estimation and spatial localization, which require more detailed structural interpretation than binary detection. While all models perform strongly for presence identification in many regions, the similarity-assisted framework delivers the most stable and balanced performance across diverse cities and rooftop conditions.

\begin{figure}[h]
\centering
\includegraphics[width=1\columnwidth]{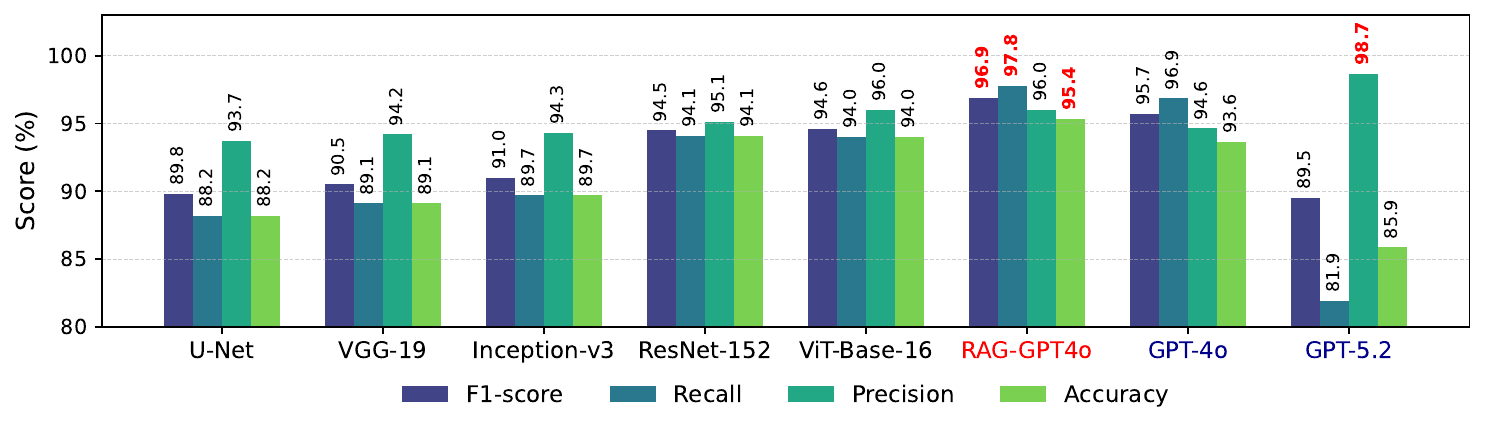}
\caption{Overall PV presence identification performance comparison. The y-axis is truncated at 80\% to emphasize performance differences between conventional computer vision architectures (U-Net to ViT) and vision-language-based approaches (GPT series). The similarity-assisted GPT-4o achieves the highest Recall and F1-score, while GPT-5.2 attains the highest Precision with reduced detection sensitivity.}
\label{fig:Overall_Presence_Performance}
\end{figure}

Fig.~\ref{fig:Overall_Presence_Performance} summarizes overall performance for PV presence identification, formulated as a binary classification problem. The results show that the similarity-assisted GPT-4o achieves the strongest overall performance, with a Recall of 97.8\% and an F1-score of 96.9\%. The increase in Recall relative to standalone GPT-4o indicates improved sensitivity to ambiguous or partially visible PV installations. Incorporation of regionally comparable reference cases appears to improve detection of subtle panel patterns without materially degrading overall reliability.

Among the conventional computer vision baselines, Vision Transformer (ViT-Base-16) and ResNet-152 achieve competitive results, with F1-scores of 94.6\% and 94.5\%, respectively. Their strong performance reflects the ability of deep feature representations to capture the geometric regularity of PV arrays under varying rooftop scales and image resolutions. In contrast, U-Net, VGG-19, and Inception-v3 exhibit lower overall performance, primarily due to reduced Recall values between 88.2\% and 89.7\%. This suggests a greater tendency toward missed detections when rooftop scenes contain visual clutter, shadowing, or diverse background textures.

A notable observation is the pronounced precision--recall trade-off exhibited by GPT-5.2. This model achieves the highest Precision at 98.7\%, indicating strong suppression of false positives. However, this occurs alongside a substantially lower Recall of 81.9\% and an overall Accuracy of 85.9\%. These results indicate a conservative detection tendency, where uncertain or ambiguous rooftop patterns are more frequently classified as non-PV. While such behavior may be appropriate in applications prioritizing false-positive avoidance, it is less suitable for comprehensive geographic PV surveys where missed installations carry greater consequences.

Overall, the results highlight the benefit of incorporating similarity-based regional reference information into the estimation process. Whereas conventional models rely solely on learned feature representations, the similarity-assisted GPT-4o integrates comparable rooftop cases during inference to refine its interpretation of ambiguous patterns. This combined strategy yields a more favorable balance between Precision and Recall, resulting in the most consistent and reliable performance for large-scale PV presence identification across diverse rooftop conditions.

{\color{black}
\subsection{Ablation Studies and Sensitivity Analysis}
\begin{figure}[htbp]
     \centering
     \begin{subfigure}[b]{0.48\textwidth}
         \centering
         \includegraphics[width=\textwidth]{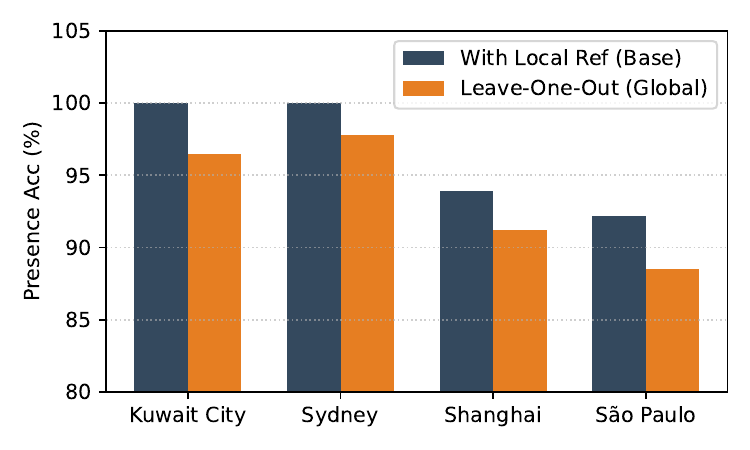}
         \caption{Cross-Regional Validation}
         \label{fig:LeaveOneOut}
     \end{subfigure}
     \hfill
     \begin{subfigure}[b]{0.48\textwidth}
         \centering
         \includegraphics[width=\textwidth]{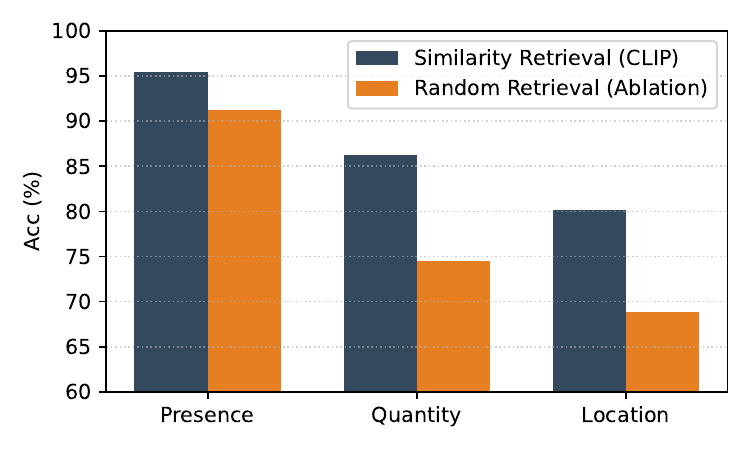}
         \caption{Random Retrieval Ablation}
         \label{fig:RandomRetrieval}
     \end{subfigure}
        \caption{Validation of Solar-RAG robustness. (a) performance remains stable even when local geographic references are excluded from the database. (b) performance significantly degrades when similarity-based retrieval is replaced with random context.}
        \label{fig:RobustnessAblation}
\end{figure}

To evaluate the geographic generalization potential of the Solar-RAG framework, we conducted a rigorous Cross-Regional Validation (Leave-One-Out) test. In this experiment, a target region (e.g., Kuwait City) is completely excluded from the retrieval database, forcing the model to rely exclusively on ``global knowledge'' from geographically distant regions to assess the held-out city. This setup simulates a real-world deployment scenario where local reference data may be unavailable. As illustrated in Fig. \ref{fig:LeaveOneOut}, removing city-specific references results in only a marginal performance dip; notably, Kuwait City and Sydney maintained high presence accuracies of 96.5\% and 97.8\%, respectively. These results confirm that the framework does not merely rely on regional memorization but instead leverages fundamental visual cues, such as rooftop geometry and climate-specific architectural patterns, to maintain high accuracy across diverse, unfamiliar geographies without requiring region-specific retraining.

Further, we conducted a Random Retrieval Ablation experiment to confirm that the observed performance gains were driven by CLIP-based similarity rather than merely the presence of additional prompt context. Comparing similarity-based retrieval against a baseline where reference images were selected randomly, as illustrated in Fig. \ref{fig:RandomRetrieval}, revealed significant performance gaps across all tasks. While PV Presence detection saw a -4.2\% decrease, the drop was far more pronounced in Quantity Estimation (-11.7\%) and Spatial Localization (-11.3\%). This steep decline confirms that VLM relies heavily on visually and contextually similar ``anchors'' to calibrate its reasoning process and produce accurate, grounded assessments.

\begin{figure}[h]
\centering
\includegraphics[width=0.7\columnwidth]{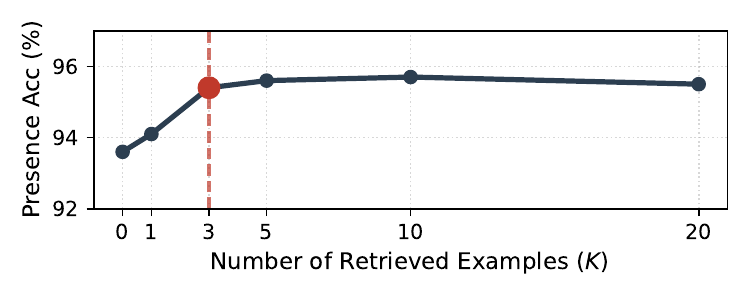}
\caption{Sensitivity analysis of the number of retrieved examples ($K$). The inflection point at $K=3$ represents the optimal balance between contextual grounding and prompt efficiency.}
\label{fig:SensitivityAnalysisK}
\end{figure}

Finally, a sensitivity analysis of $K$ was performed to justify the selection of $K=3$ as the optimal retrieval volume for the framework. The results in Fig. \ref{fig:SensitivityAnalysisK} indicate a sharp increase in Presence Detection Accuracy as $K$ moves from 0 (93.6\%) to 3 (95.4\%), representing the most significant efficiency gain. Beyond this point, the accuracy curve flattens significantly, with $K=5$ and $K=10$ yielding negligible improvements. This ``elbow point'' at $K=3$ suggests that a small, high-quality set of visually similar reference scenes provides sufficient contextual grounding for robust assessment while avoiding the noise or context-window saturation that can occur with larger values of $K$.
}

\section{Feeder-Level Power System Impact Analysis}
\label{sec:ps-coupling}
While the previous sections evaluate the proposed PV assessment framework at the image and city levels, practical power-system applications ultimately depend on how rooftop PV estimation errors affect feeder-level quantities such as net load and bus voltage magnitudes. This section presents a stylized feeder simulation to illustrate how the structured PV outputs defined in Section~\ref{sec: Problem Formulation} can be integrated into conventional AC power-flow analysis tools. The objective is to connect image-level structured predictions with distribution-level system evaluation. Such coupling analysis enables evaluation of how upstream PV inventory uncertainty may influence downstream distribution planning studies, including net-load estimation and voltage regulation analysis.

To illustrate this coupling mechanism, we build upon the standard 30-bus AC test system implemented in PYPOWER. A full 24-hour period is simulated with 15-minute resolution. Rooftop PV capacities derived from image-level assessments are mapped to bus-level active power injections, and for each PV assessment method we compute the corresponding time-varying feeder net-load trajectory and bus voltage profile. This enables direct comparison of system-level deviations resulting from different PV estimation approaches.

\subsection{Temporal Discretization, Load, and PV Profiles}
\label{subsec:time-profiles}

We consider one representative operating day with a fixed time step
\begin{equation}
  \Delta t = 15~\mathrm{min}.
\end{equation}
Let $t \in \{1,\dots,T\}$ index the discrete time steps, where $T = 96$, and denote the corresponding continuous time in hours by
\begin{equation}
  \tau_t = (t-1)\Delta t, 
  \qquad \tau_t \in [0,24).
\end{equation}

Let $P_{d,i}^0$ and $Q_{d,i}^0$ denote the nominal active and reactive demands (in MW and MVAr) at bus $i$ in the base case of the 30-bus system. We construct a normalized diurnal load profile $\ell(\tau) \in [0,1]$ with a morning increase, a dominant mid-day peak, and an evening peak, reflecting typical residential or mixed-use summer demand behavior. The time-varying demands are given by
\begin{equation}
  P_{d,i}(t) = \alpha_L\,\ell(\tau_t)\,P_{d,i}^0, 
  \qquad
  Q_{d,i}(t) = \alpha_L\,\ell(\tau_t)\,Q_{d,i}^0,
  \label{eq:time-varying-load}
\end{equation}
where $\alpha_L > 1$ is a load scaling factor used to adjust overall feeder loading conditions. 
In parallel, we define a normalized PV generation profile $g(\tau) \in [0,1]$ representing clear-sky rooftop PV production between 06{:}00 and 18{:}00. The profile increases after sunrise, reaches its maximum near solar noon, and decreases toward sunset.  
Let $\tilde{C}_i^{(m)}$ denote the scaled PV capacity assigned to bus $i$ under assessment model $m$ (as defined in Section~\ref{subsec:pv-mapping}). The instantaneous PV active power injection at bus $i$ and time $t$ is proportional to the generation profile and installed capacity, i.e.,
\begin{equation}
  P_{\mathrm{PV},i}^{(m)}(t)
  = g(\tau_t)\,\tilde{C}_i^{(m)}.
\end{equation}

\subsection{From Image-Level Assessments to Bus-Level PV Capacity}
\label{subsec:pv-mapping}

As formulated in Section~\ref{sec: Problem Formulation}, the proposed PV assessment framework produces, for each satellite image, a structured output
\begin{equation}
\mathbf{z}^{\mathrm{PV}} = \left[ p^{\mathrm{PV}}, n^{\mathrm{PV}}, \ell^{\mathrm{PV}}, e^{\mathrm{PV}} \right],
\end{equation}
where $p^{\mathrm{PV}}$ indicates PV presence, $n^{\mathrm{PV}}$ denotes the panel-quantity interval, and $\ell^{\mathrm{PV}}$ specifies a coarse spatial location within the image.
For feeder-level simulation, we utilize $p^{\mathrm{PV}}$, $n^{\mathrm{PV}}$, and $\ell^{\mathrm{PV}}$ to determine PV capacities and their corresponding bus assignments.
We consider a collection of rooftop sites indexed by $r$, each electrically connected to a feeder bus $b_r$. For each site $r$, the \emph{true} panel-quantity interval is denoted
\begin{equation}
q_r^{\mathrm{true}} \in \mathcal{Q}
= \{(0,1], (1,5], (5,10], (10,\infty)\},
\label{eq:q-intervals}
\end{equation}
consistent with the discrete interval labels used in the PV assessment task. Each interval $q \in \mathcal{Q}$ is mapped to a representative panel count $n(q)$, for example
\begin{equation}
n((0,1]) = 0.5,\quad n((1,5]) = 3,\quad
n((5,10]) = 7,\quad n((10,\infty)) = 12.
\label{eq:interval-to-panels}
\end{equation}

Assuming a per-panel rating of $P_{\text{panel}} = 0.4~\mathrm{kW}$, the true installed PV capacity at rooftop site $r$ (in kW) is
\begin{equation}
C_r^{\mathrm{true}} = n\left(q_r^{\mathrm{true}}\right) P_{\text{panel}}.
\label{eq:site-true-capacity}
\end{equation}

For each assessment model $m \in \{\mathrm{RAG}, \mathrm{4o}, \mathrm{5.2}\}$, we construct corresponding model-specific quantity intervals $q_r^{(m)}$ and bus assignments $b_r^{(m)}$. These are generated from the true quantities and locations using an error-generation procedure calibrated to the empirical panel-quantity and spatial-localization accuracies reported in Table~\ref{tab:results_citywise}.
Specifically, for each site $r$:
(i) with probability $a_q^{(m)}$ (the panel-quantity accuracy of model $m$), we set $q_r^{(m)} = q_r^{\mathrm{true}}$; otherwise, $q_r^{(m)}$ is assigned to a neighboring interval in $\mathcal{Q}$ according to a predefined misclassification pattern reflecting typical under- or over-estimation;
(ii) with probability $a_\ell^{(m)}$ (the spatial-localization accuracy of model $m$), we set $b_r^{(m)} = b_r$; otherwise, $b_r^{(m)}$ is assigned to an electrically adjacent bus in the feeder topology.
Given the model-specific interval $q_r^{(m)}$, the estimated PV capacity at site $r$ (in kW) under model $m$ is
\begin{equation}
C_r^{(m)} = n\left(q_r^{(m)}\right) P_{\text{panel}},
\label{eq:site-capacity-m}
\end{equation}
and the aggregate PV capacity connected to bus $i$ (in kW) becomes
\begin{equation}
C_i^{(m)} = \sum_{r:, b_r^{(m)} = i} C_r^{(m)}.
\label{eq:bus-capacity-m}
\end{equation}

To represent a high-penetration rooftop PV scenario, all bus-level capacities are scaled by a common factor $\kappa$ such that, in the true-PV case, rooftop PV supplies a fixed fraction $\rho$ of the system peak load:
\begin{equation}
\kappa =
\frac{\rho \sum_i P_{d,i}^0 \alpha_L}
    {\sum_i C_i^{\mathrm{true}}},
\qquad
\tilde{C}_i^{(m)} = \kappa C_i^{(m)}.
\label{eq:pv-scaling-factor}
\end{equation}
Here $\tilde{C}_i^{(m)}$ denotes the scaled installed PV capacity at bus $i$ under model $m$.
The active PV injection at bus $i$ and time step $t$ under model $m$ (in MW) is then
\begin{equation}
  P_{\mathrm{PV},i}^{(m)}(t)
  = g(\tau_t)\,\tilde{C}_i^{(m)} / 1000,
  \label{eq:pv-injection}
\end{equation}
where $g(\tau_t)$ denotes the normalized clear-sky PV generation profile at time $t$, and the division by $1000$ converts kW to MW.
\subsection{AC Power Flow Model and Net-Load Trajectories}
\label{subsec:acpf}

For each assessment model $m$ and time step $t$, the net active and reactive demands supplied to the AC power-flow model are defined as
\begin{equation}
  \hat{P}_{d,i}^{(m)}(t)
  = P_{d,i}(t) - P_{\mathrm{PV},i}^{(m)}(t),
  \qquad
  \hat{Q}_{d,i}^{(m)}(t)
  = Q_{d,i}(t),
  \label{eq:net-load-bus}
\end{equation}
where PV injections reduce the effective active demand at each bus.
These quantities, together with the fixed network parameters of the 30-bus system, define a standard nonlinear AC load-flow problem, which is solved using the PYPOWER solver for each pair $(m,t)$.
Let $V_i^{(m)}(t)$ denote the voltage magnitude at bus $i$, and let $P_{\mathrm{sub}}^{(m)}(t)$ denote the active power injection at the slack generator (interpreted as feeder-level net load) at time $t$ under model $m$. 
The reference (true-PV) case corresponds to $m = \mathrm{true}$, where the unperturbed capacities $C_i^{\mathrm{true}}$ and original bus assignments $b_r$ are used.

\subsection{Error Metrics for Power-System Coupling}
\label{subsec:metrics}

The feeder simulation quantifies how inaccuracies in rooftop PV assessment propagate to feeder-level quantities relevant to distribution system operation and planning. In this study, we focus on two primary aspects:

\begin{enumerate}
  \item the \emph{installed PV capacity} observed at the feeder level, obtained by aggregating bus-level PV capacities; and
  \item the \emph{time-varying feeder net load} at the substation over a full day with 15-minute resolution.
\end{enumerate}

For each assessment model $m \in \{\mathrm{RAG}, \mathrm{4o}, \mathrm{5.2}\}$, let $\tilde{C}_i^{(m)}$ denote the scaled PV capacity (in kW) assigned to bus $i$ in~\eqref{eq:pv-scaling-factor}. The total installed rooftop PV capacity connected to the feeder under model $m$ is therefore
\begin{equation}
  C_{\mathrm{tot}}^{(m)} = \sum_i \tilde{C}_i^{(m)} ,
  \label{eq:total-capacity}
\end{equation}
where $C_{\mathrm{tot}}^{(m)}$ is expressed in kW. For reporting purposes, this quantity is converted to MW when presented in tables. We compare $C_{\mathrm{tot}}^{(m)}$ with the true-PV reference value $C_{\mathrm{tot}}^{\mathrm{true}}$ to assess systematic over- or under-estimation of installed rooftop PV capacity. 
Let $P_{\mathrm{sub}}^{\mathrm{true}}(t)$ denote the feeder net load (slack-generator active power injection) in the reference case $m=\mathrm{true}$ at time step $t$, and let $P_{\mathrm{sub}}^{(m)}(t)$ denote the corresponding value under assessment model $m$. The 24-hour root-mean-square error (RMSE) of feeder net load is defined as
\begin{equation}
  \mathrm{RMSE}^{(m)} =
  \sqrt{\frac{1}{T}\sum_{t=1}^T
  \bigl(P_{\mathrm{sub}}^{(m)}(t) -
        P_{\mathrm{sub}}^{\mathrm{true}}(t)\bigr)^2 } ,
  \label{eq:rmse}
\end{equation}
where $\mathrm{RMSE}^{(m)}$ is expressed in MW and summarizes the average deviation of the estimated net-load trajectory from the true trajectory over the full operating day. 
To provide a relative error measure, we additionally compute the mean absolute percentage error (MAPE). Let
\begin{equation}
  \mathcal{T}_{\mathrm{nz}}
  = \left\{ t :
  \left| P_{\mathrm{sub}}^{\mathrm{true}}(t) \right| \ge \varepsilon \right\}
\end{equation}
denote the set of time steps for which the absolute true net load exceeds a small threshold $\varepsilon > 0$ (e.g., 5\% of the maximum absolute net load), thereby avoiding numerical inflation of relative errors when the denominator is close to zero. The MAPE of feeder net load is defined as
\begin{equation}
  \mathrm{MAPE}^{(m)} =
  \frac{100}{|\mathcal{T}_{\mathrm{nz}}|}
  \sum_{t \in \mathcal{T}_{\mathrm{nz}}}
  \left|
    \frac{P_{\mathrm{sub}}^{(m)}(t) -
          P_{\mathrm{sub}}^{\mathrm{true}}(t)}
         {P_{\mathrm{sub}}^{\mathrm{true}}(t)}
  \right|.
  \label{eq:mape}
\end{equation}

In the results section, we present:
(i) the time series of $P_{\mathrm{sub}}^{(m)}(t)$ over the full 24-hour period,
(ii) the aggregated capacities $C_{\mathrm{tot}}^{(m)}$, and
(iii) the scalar metrics $\mathrm{RMSE}^{(m)}$ and $\mathrm{MAPE}^{(m)}$ for each assessment model.

To examine spatial effects, we additionally visualize heatmaps of the absolute voltage magnitude deviation
\begin{equation}
  E_{V,i}^{(m)}(t)
  = \left| V_i^{(m)}(t) - V_i^{\mathrm{true}}(t) \right|
  \label{eq:voltage-error}
\end{equation}
across all buses $i$ and time steps $t$. This provides a bus-time representation of how rooftop PV estimation errors translate into voltage deviations relative to the true-PV reference case. All feeder-level results are obtained using the same calibrated quantity and localization error assumptions derived from the image-level evaluation.

\subsection{Power-System Results, Discussion, and Limitations}
\label{subsec:ps-results}

\begin{table}[t]
  \centering
  \caption{Feeder-level PV capacity and 24-hour net-load error metrics for different assessment models on the PYPOWER 30-bus system. The true-PV case ($m=\mathrm{true}$) is used as the reference for all error computations.}
  \label{tab:ps-metrics}
  \begin{tabular}{lccc}
    \hline
    Model &
    $C_{\mathrm{tot}}$ [MW] &
    RMSE [MW] &
    MAPE [\%] \\
    \hline
    True PV ($m=\mathrm{true}$)      & 147.58 & --    & --     \\
    RAG--GPT-4o ($m=\mathrm{RAG}$)   & 146.81 & 0.34  & 0.44   \\
    GPT-4o ($m=\mathrm{4o}$)         & 142.22 & 2.37  & 3.13   \\
    GPT-5.2 ($m=\mathrm{5.2}$)       & 115.27 & 15.35 & 19.43  \\
    \hline
  \end{tabular}
\end{table}

\begin{figure}[h]
\centering
\includegraphics[width=0.6\columnwidth]{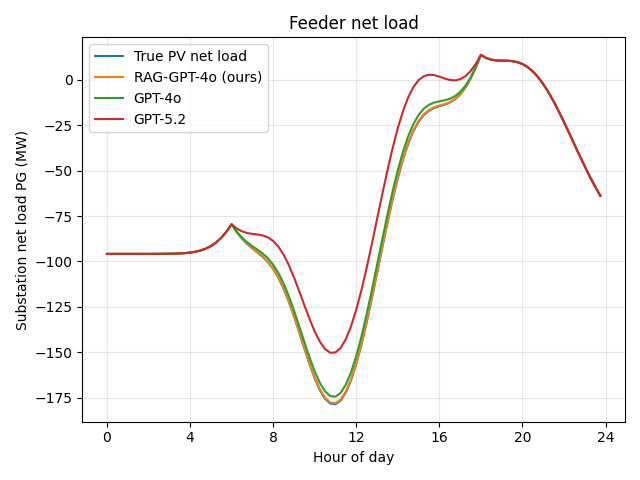}
\caption{Feeder-level net-load trajectories $P_{\mathrm{sub}}^{(m)}(t)$ under different assessment models. The curves show the substation active power injection over a 24-hour period with 15-minute resolution for the reference case ($m=\mathrm{true}$) and for PV capacities derived from $m\in\{\mathrm{RAG},\mathrm{4o},\mathrm{5.2}\}$. All cases share identical time-varying load and PV generation profiles on the PYPOWER 30-bus system. The RAG-enhanced model produces a net-load trajectory closely matching the reference case, whereas GPT-4o moderately and GPT-5.2 significantly overestimate net load during high-PV hours due to underestimation of rooftop PV capacity.}
\label{fig:ps-netload}
\end{figure}

\begin{figure}[h]
\centering
\includegraphics[width=1\columnwidth]{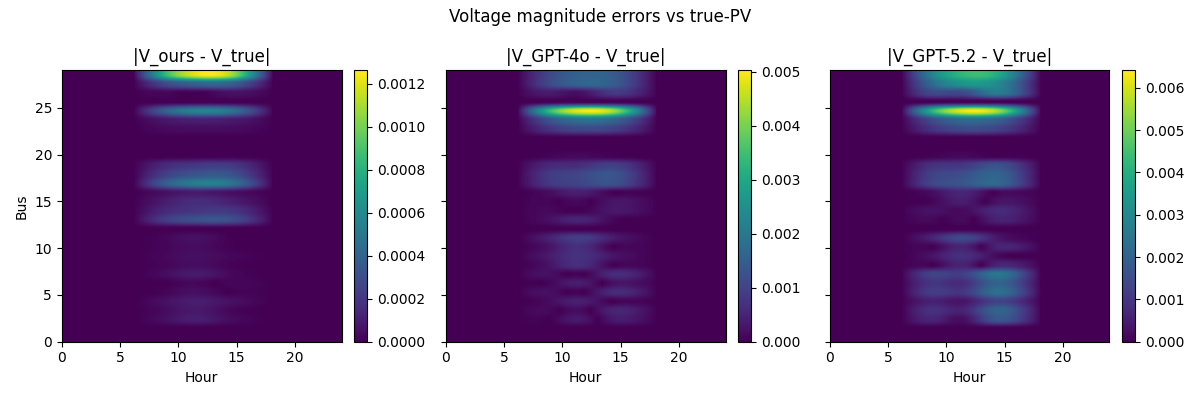}
\caption{Voltage magnitude deviation heatmaps relative to the reference case. Each panel shows $E_{V,i}^{(m)}(t) = \left| V_i^{(m)}(t) - V_i^{\mathrm{true}}(t) \right|$ for all buses (vertical axis) and 15-minute time steps (horizontal axis) in the PYPOWER 30-bus feeder. The RAG-enhanced model yields small, localized voltage deviations on the order of $10^{-3}$~p.u. GPT-4o produces larger deviations affecting more buses, and GPT-5.2 results in the largest and most widespread voltage magnitude errors, particularly during midday high-PV periods.}
\label{fig:ps-vheatmap}
\end{figure}

Fig.~\ref{fig:ps-netload} and Fig.~\ref{fig:ps-vheatmap} illustrate how rooftop PV assessment errors propagate to feeder-level quantities in the 30-bus AC test system. The structured image-level outputs of assessment models $m \in \{\mathrm{RAG}, \mathrm{4o}, \mathrm{5.2}\}$ are mapped to bus-level PV capacities as described in Section~\ref{sec:ps-coupling}, and a full 24-hour simulation is performed with 15-minute resolution using identical load and PV generation profiles.

\paragraph{Installed PV capacity}

Table~\ref{tab:ps-metrics} compares the total rooftop PV capacity $C_{\mathrm{tot}}^{(m)}$ connected to the feeder under each assessment model. In the reference case ($m=\mathrm{true}$), the scaled capacity is 
\[
C_{\mathrm{tot}}^{\mathrm{true}} = 147.58~\mathrm{MW}.
\]
The RAG-enhanced model ($m=\mathrm{RAG}$) estimates 
\[
C_{\mathrm{tot}}^{(\mathrm{RAG})} = 146.81~\mathrm{MW},
\]
corresponding to a relative error of $-0.52\%$. GPT-4o ($m=\mathrm{4o}$) underestimates capacity by $-3.6\%$, while GPT-5.2 ($m=\mathrm{5.2}$) exhibits a substantial underestimation of $-21.9\%$. 
These feeder-level biases directly reflect image-level quantity estimation accuracy. Even moderate misclassification of panel-count intervals accumulates into meaningful capacity distortion when aggregated across many rooftop sites.

\paragraph{Feeder net-load trajectories}

Fig.~\ref{fig:ps-netload} shows the resulting substation net load $P_{\mathrm{sub}}^{(m)}(t)$ over 24 hours. During night-time periods, when PV generation is negligible, all trajectories align closely. As solar output increases, deviations emerge. GPT-5.2 significantly overestimates net load during peak solar hours due to underestimation of installed PV capacity. GPT-4o shows smaller but noticeable deviations. The RAG-enhanced model remains closely aligned with the reference trajectory throughout the day. 
The quantitative impact is summarized by the metrics $\mathrm{RMSE}^{(m)}$ and $\mathrm{MAPE}^{(m)}$ in Table~\ref{tab:ps-metrics}. The RAG-enhanced model achieves 
\[
\mathrm{RMSE}^{(\mathrm{RAG})} = 0.34~\mathrm{MW},
\]
compared to $2.37~\mathrm{MW}$ for $m=\mathrm{4o}$ and $15.35~\mathrm{MW}$ for $m=\mathrm{5.2}$. Relative errors follow the same pattern, with
\[
\mathrm{MAPE}^{(\mathrm{RAG})} = 0.44\%,\quad
\mathrm{MAPE}^{(\mathrm{4o})} = 3.13\%,\quad
\mathrm{MAPE}^{(\mathrm{5.2})} = 19.43\%.
\]
These results demonstrate that systematic PV underestimation can induce significant feeder-level net-load distortion.

\paragraph{Voltage magnitude deviations}

Voltage deviation heatmaps in Fig.~\ref{fig:ps-vheatmap} provide a spatial perspective. The RAG-enhanced model produces small and localized deviations $E_{V,i}^{(\mathrm{RAG})}(t)$ (on the order of $10^{-3}$~p.u.), primarily during peak solar hours. GPT-4o introduces larger deviations across more buses. GPT-5.2 results in the most widespread voltage errors, particularly at downstream buses during high-generation periods.

\paragraph{Discussion and limitations}

These results demonstrate that improvements in image-level PV quantity and localization accuracy translate into materially reduced feeder-level capacity bias, net-load distortion, and voltage deviation. The RAG-enhanced framework improves numerical reliability without modifying the underlying AC power-flow formulation, highlighting the importance of contextual grounding in upstream PV assessment.

Several limitations should be acknowledged. First, the framework provides interval-based panel quantity estimates and coarse spatial localization rather than pixel-level segmentation, which may limit applications requiring precise capacity calculation or geometric modeling. Second, performance depends on the diversity and representativeness of the reference database; rare rooftop configurations or underrepresented regions may reduce retrieval effectiveness. Third, the feeder simulation relies on a stylized test system and probabilistic error modeling calibrated from image-level metrics, rather than a real utility feeder with measured PV capacities.

Future work will explore finer-grained spatial modeling, expanded multi-region reference databases, integration with real feeder datasets, and computational optimizations for large-scale deployment. Extending the framework to jointly estimate PV orientation, tilt, and shading effects may further improve capacity modeling and downstream distribution-level analysis.

\section{Conclusion}
\label{sec:Conclusion}
This paper presented a retrieval-augmented framework for satellite-based rooftop photovoltaic (PV) inventory estimation using geo-referenced imagery. By incorporating visually similar reference rooftop scenes into multimodal vision--language reasoning, the proposed Solar-RAG approach improves robustness to geographic variability and produces structured PV descriptors, including installation presence, panel quantity intervals, and coarse spatial location. These structured outputs enable scalable estimation of distributed PV deployment across diverse rooftop environments without requiring region-specific model retraining.
Extensive cross-regional experiments demonstrate that similarity-based reference grounding systematically improves PV assessment accuracy compared with standalone vision--language models, particularly for panel quantity estimation and spatial localization. The retrieval-assisted framework also maintains a favorable precision-recall balance for PV presence detection across geographically heterogeneous datasets.

Beyond image-level evaluation, feeder-level simulations using a benchmark AC power-flow model show that improved PV quantity and localization accuracy significantly reduce errors in aggregated PV capacity, feeder net-load trajectories, and bus voltage magnitudes. These results demonstrate how structured rooftop PV inventory estimates derived from satellite imagery can be directly incorporated into distribution-system analysis and planning workflows.
Overall, the study highlights the potential of retrieval-assisted multimodal reasoning as a practical approach for large-scale PV inventory estimation and distributed generation monitoring. Future work will focus on integrating the framework with real feeder datasets, expanding geographically diverse reference repositories, and extending the methodology to estimate additional PV system characteristics such as orientation, tilt, and shading effects for enhanced power-system modeling.

\bibliographystyle{elsarticle-num} 
\bibliography{Bibliography}
\end{document}